\documentclass[twocolumn,showpacs,amsmath,amssymb,aps,prb,superscriptaddress]{revtex4-1}

\usepackage[]{graphicx}
\usepackage[]{verbatim}
\usepackage[]{color}
\usepackage{overpic}
\usepackage{rotating}
\usepackage{mathtools}
\usepackage{appendix}
\usepackage{ulem}
\usepackage[margin=0.6in]{geometry}

\usepackage[]{hyperref}

 \hypersetup{
 colorlinks = true,
 linkcolor = blue,
 urlcolor = magenta,
 citecolor = magenta}

\usepackage{array}
\newcolumntype{L}[1]{>{\raggedright\let\newline\\\arraybackslash\hspace{0pt}}m{#1}}
\newcolumntype{C}[1]{>{\centering\let\newline\\\arraybackslash\hspace{0pt}}m{#1}}
\newcolumntype{R}[1]{>{\raggedleft\let\newline\\\arraybackslash\hspace{0pt}}m{#1}}
\newcolumntype{N}{@{}m{0pt}@{}}

\makeatletter
\newsavebox{\@brx}

\newcommand{\llangle}[1][]{\savebox{\@brx}{\(\m@th{#1\langle}\)}%
  \mathopen{\copy\@brx\mkern2mu\kern-0.8\wd\@brx\usebox{\@brx}}}
\newcommand{\rrangle}[1][]{\savebox{\@brx}{\(\m@th{#1\rangle}\)}%
  \mathclose{\copy\@brx\mkern2mu\kern-0.8\wd\@brx\usebox{\@brx}}}

  \newcommand{\lllangle}[1][]{\savebox{\@brx}{\(\m@th{#1\langle}\)}%
  \mathopen{\copy\@brx\copy\@brx\mkern4mu\kern-0.7\wd\@brx\usebox{\@brx}}}
\newcommand{\rrrangle}[1][]{\savebox{\@brx}{\(\m@th{#1\rangle}\)}%
  \mathclose{\copy\@brx\copy\@brx\mkern4mu\kern-0.7\wd\@brx\usebox{\@brx}}}

\graphicspath{{./}{./}}

\begin{document}
\title{Counter-rotating spiral order in three dimensional iridates: signature of hidden symmetry in Kitaev-$\Gamma$ model}
\author{P. P. Stavropoulos}
\affiliation{Department of Physics and Center for Quantum Materials, University of Toronto, 60 St.~George St., Toronto, Ontario, M5S 1A7, Canada}
\author{Andrei Catuneanu}
\affiliation{Department of Physics and Center for Quantum Materials, University of Toronto, 60 St.~George St., Toronto, Ontario, M5S 1A7, Canada}
\author{Hae-Young Kee}
\affiliation{Department of Physics and Center for Quantum Materials, University of Toronto, 60 St.~George St., Toronto, Ontario, M5S 1A7, Canada}
\affiliation{Canadian Institute for Advanced Research, Toronto, Ontario, M5G 1Z8, Canada}
\email{hykee@physics.utoronto.ca}

\begin{abstract}
The unconventional magnetic orderings found in Kitaev spin liquid candidates
suggest frustration of spin interactions, and raise a possibility of nearby spin liquid phases.
In particular, counter-rotating spiral ordering in three-dimensional (3D)  iridates is striking, and
understanding the microscopic mechanism of such ordering may provide routes to 3D Kitaev spin liquids. 
 We study a minimal 3D model including Kitaev $K$ and a symmetric off-diagonal bond-dependent
 $\Gamma$ interaction on the hyperhoneycomb lattice using exact diagonalization.
 We first show that a 12-site unitary transformation unveils a Heisenberg model with hidden SU(2) symmetry when $K=\Gamma$.
 The magnetic ordering in the transformed basis then generically maps to the counter-rotating noncoplanar spiral order of the original spin.
 The moment direction depends on perturbations away from the SU(2) point. When $K$ and $\Gamma$ are negative,
 a positive Heisenberg interaction favors the $(110)$-direction reported in the neutron scattering measurement on $\beta$-Li$_2$IrO$_3$. 
Our findings offer a relevant set of microscopic parameters, which in turn guides
  a way to approach possible Kitaev spin liquids.
 \end{abstract}
\maketitle

\section{introduction}
Quantum spin liquids are one of most fascinating novel phases of matter, possessing topological order and fractionalized excitations.\cite{AndersonMRB1973, BalentsNat2010}
Since the Heisenberg interaction favors either antiferromagnetic or ferromagnetic ordering on a bipartite lattice, it was suggested that geometrical frustration, such as on the Kagome lattice, is required to achieve a spin liquid phase. 
Kitaev found that bond dependent interactions are another route to introduce spin frustration and constructed an exactly solvable Z$_2$ spin liquid model on the two-dimensional honeycomb lattice.\cite{KitaevAP2006} 
However it was not clear how to generate such a bond dependent interaction in solid-state materials.
Jackeli and Khaliullin showed that strong spin-orbit coupling together with strong electronic interactions leads to such an interaction on honeycomb lattices with edge sharing octahedra.\cite{JackeliPRL2009}

Since then, there has been an active search for Kitaev spin liquid candidates.
So far they include two-dimensional (2D) layered honeycomb iridates\cite{SinghPRB2010, GegenwartPRL2012}, $\alpha$-RuCl$_3$\cite{KimPRB2014,KeePRB2015}, and three-dimensional (3D) hyper- and harmonic-honeycomb iridates \cite{AnalytisNC2014,TakagiPRL2015}.
%
At low temperatures, these candidates order magnetically instead of becoming spin liquids, but the magnetic ordering patterns are unconventional. 
Na$_2$IrO$_3$\cite{HillPRB2011, TaylorPRL2012, CaoPRB2012} and $\alpha$-RuCl$_3$\cite{KimPRB2015}
exhibit a zig-zag ordering, while Li$_2$IrO$_3$ and 3D iridates display an incommensurate counter-rotating spiral ordering.\cite{ColdeaPRB2014, ColdeaPRL2014,ColdeaPRB2016}
The magnetic order suggests that other spin interactions apart from the Kitaev term must play a role, and thus the 
Kitaev-Heisenberg ($K$H) model was proposed to explain the reported zig-zag magnetic ordering.\cite{KhaliullinPRL2010,TrebstPRB2011, 
ChaloupkaPRL2013, 
PerkinsPRB2014}

A generic nearest neighbour spin model on an ideal honeycomb lattice derived by Rau et al\cite{RauPRL2014}
found a bond dependent symmetric off-diagonal $\Gamma$-term 
and a complete Kitaev-$\Gamma$-Heisenberg ($K\Gamma$H) model was obtained.
Roughly,  $\Gamma \propto  \frac{J_H t_d t_o}{U^2}$, while $K \propto  \frac{J_H t_o^2}{U^2}$
and $ J \propto \frac{t_d^{'2}}{U}$, where $J_H$, $t_d/t_d^{'}$, $t_o$ and $U$ are the Hund's coupling, 
 effective direct-, indirect-hopping integrals, and Hubbard interaction, respectively.
When there are only indirect hoppings, it reduces to the Kitaev model.
 The presence of the $\Gamma$ interaction was confirmed by \textit{ab initio} \cite{ImadaPRL2014}, and quantum molecular calculations\cite{VanderBrinkNJoP2014}.
 Since the spin interactions in general depend on hopping parameters, spin-orbit coupling, and microscopic Coulomb interactions, 
 their strengths vary among different materials.\cite{KrempaARCMP2014,RauARCMP2016}
 \textit{Ab initio} studies and quantum molecular calculations on 3D iridates \cite{VanDenBrinkSR2016, KimEPL2015, KimPRB2016} reported different values, and both $K$ and $\Gamma$ are negative and dominate over the Heisenberg term.\cite{KimEPL2015, KimPRB2016}

The classical $K\Gamma$H model was studied for 3D systems in Ref. \citenum{LeePRB2015} and several ordered phases were found.
Among them, the SP$_{a^-}$ phase has the same symmetry as the experimentally observed magnetic ordering pattern, as determined by neutron scattering.\cite{ColdeaPRB2014}
This phase occurs when  $K$ and $\Gamma$ are negative with a small positive Heisenberg interaction.
It is worthwhile to note that the Kitaev-$\Gamma$ (K$\Gamma$) model has a hidden SU(2) symmetry via a 6-site transformation when $K= \Gamma$ in the 2D honeycomb lattice model.
This point in phase space corresponds to the ordered phase denoted as the 120$^\circ$ state\cite{RauPRL2014} or vortex state\cite{KhaliullinPRB2015}.
Another classical spin model study found two different phases near the same parameter space depending on the relative strength of $K$ and $\Gamma$ .\cite{NataliaPRB2018}

Identifying spin liquids and nearby phases in quantum spin models in 3D is a challenging task. 
Exact diagonalization (ED) is difficult to perform due to the large Hilbert space. 
However, if there is an exactly solvable point in the K$\Gamma$ model, one may limit the parameter space
around the solvable point and search for possible ordered phases and spin liquids nearby.
The symmetric point suggests a natural way to construct a suitable ED cluster, as it may involve a finite number of sites
similar to the 4- or 6-site transformation found in the 2D honeycomb lattice.\cite{KhaliullinPRB2015}

Here we study a 3D $K\Gamma$H model using ED method. We first search for an exactly solvable point in the 3D hyperhoneycomb lattice.
We derive a Heisenberg model with a hidden SU(2) symmetry through a 12-site transformation when $K = \Gamma$. 
While our main focus is on the hyperhoneycomb, it can be generalized to other lattices with the same local geometry, i.e. 
three ($X$, $Y$, and $Z$) bond-dependent interactions. 
Then we perform 24- and 30-site ED calculations on the $K\Gamma$H model in the regime where $K \sim \Gamma$ on the hyperhoneycomb lattice.

In particular, when  $K$ and $\Gamma$ are negative, the corresponding phase has the same symmetry as the counter-rotating spiral order observed in $\beta$-Li$_2$IrO$_3$, provided that the moment is pinned along a certain direction. 
While the ordering wavevector $Q=(2/3, 0,0)$ at the SU(2) point is commensurate, the wavevector itself may shift when Heisenberg interactions are added without losing the symmetry of the counter-rotating spiral, i.e. a feature of the underlying spin frustration.
This ordered phase occupies a wide parameter space which includes the SU(2) point, 
and the direction of moment depends on perturbations away from the SU(2) point.
The moment direction found in neutron scattering\cite{ColdeaPRB2014} is chosen when a positive Heisenberg interaction is added, and there are six equivalent moment directions related by C$_3$ rotation around the $(-1,1,1)$ axis.
We also present the evolution of this phase from the 1D (chain) to the 3D limit.
Another type of spiral ordered phase and the Kitaev liquid phases all exist in the 1D and persist into the 3D limit, and
a new phase emerges only when chains are coupled via the $Z$-bond.

Below in Sec. \ref{setup_and_su2_points} we will first show the hidden SU(2) symmetry of the $K\Gamma$ model using a 12-site transformation and 
 the corresponding counter-rotating spiral ordering pattern. 
In Sec. \ref{phasepinning}, we present 24-site ED results and examine the magnetic moment directions when the system is perturbed away from the SU(2) point, such as by adding a Heisenberg interaction.
In Sec. \ref{chains}, we take a limit of the 1D $K\Gamma$ model where an additional hidden SU(2) symmetry occurs via a 6-site transformation along the chain, and discuss the connection between the1D and 3D limits by increasing the interchain coupling. 
In the last section we summarize our findings and discuss how to approach spin liquids close by these ordered phases.

\section{SU(2) Hidden symmetry: 12-site transformation} \label{setup_and_su2_points}

The hyperhoneycomb $\beta-$Li$_2$IrO$_3$, whose conventional unit cell is shown in Fig. \ref{fig:coorddef}, can be viewed as the 3D generalization to the familiar 2D honeycomb structure.
%
The iridium sites form a tri-coordinated lattice with three bond types labeled by $X$, $Y$ and $Z$, similar to the 2D honeycomb lattice .\cite{RauPRL2014}
The different color sites (yellow and blue) which form chains in Fig. \ref{fig:coorddef} denote the sign convention of the spin interactions on the $X$- and $Y$-bonds, where different colored chains are connected by the $Z$-bonds. 

A combination of crystal field spitting due to the octahedral cages and a strong spin-orbit coupling leads to a pseudospin J$_{\rm eff}=1/2$ as the relevant spin degree of freedom in iridium oxides.\cite{RotenbergPRL2008,GretarssonPRL2013}
Furthermore, the edges of the octahedral cages are shared between two Ir atoms which results in dominant bond-dependent interactions in the large Hubbard U limit.


\subsection*{Minimal spin model}
\begin{figure}[!ht]
  \centering
  \begin{overpic}[scale=0.35]{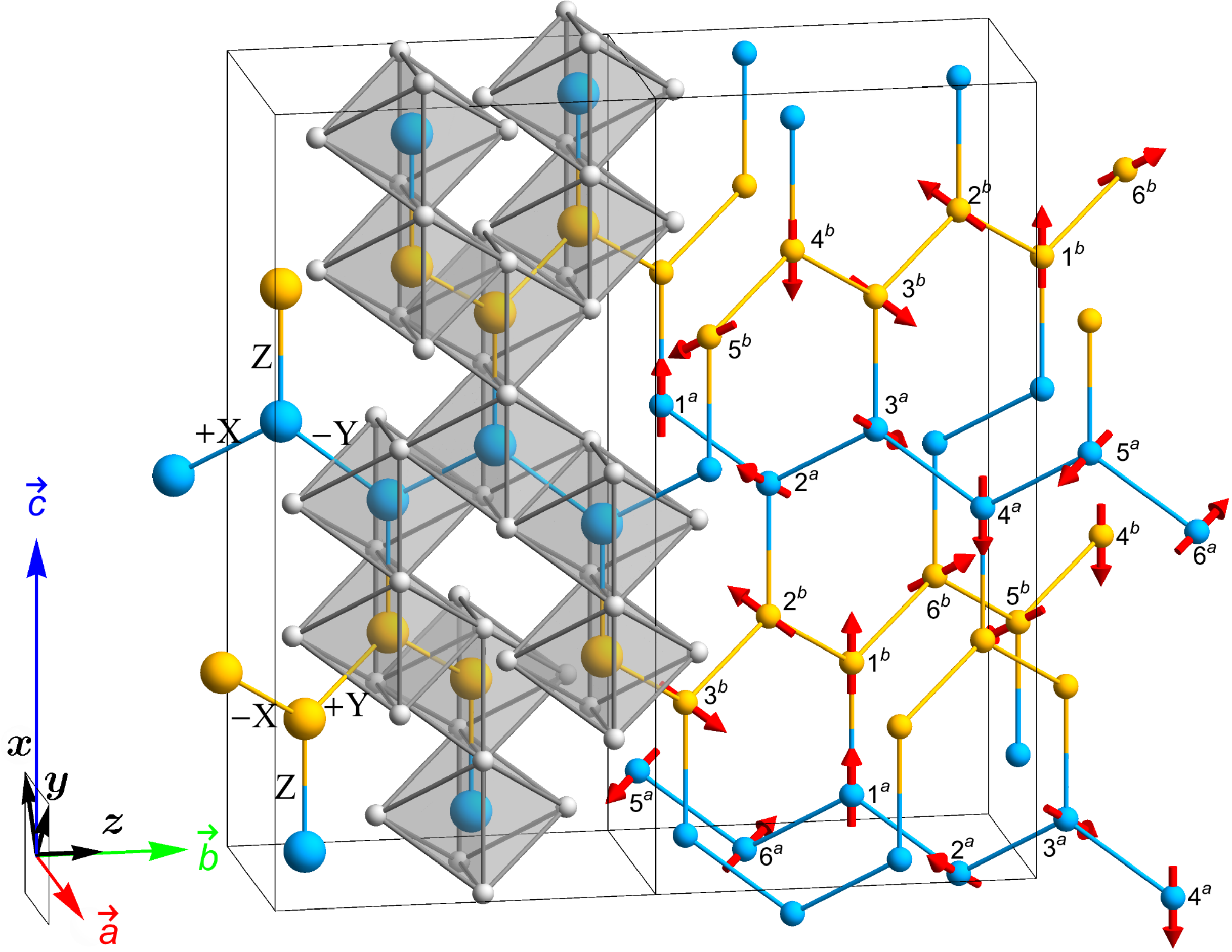}
  \end{overpic}  
  \caption{Two conventional unit cells of the ideal hyperhoneycomb structure are shown.
The lattice vectors $\vec{a}, \vec{b}, \vec{c}$ for the conventional unit cell, and $(x,y,z)$ are depicted by arrows in the left corner where $\vec{a}=(-2,2,0)$, $\vec{b}=(0,0,4)$, $\vec{c}=(6,6,0)$ in the global coordinates $(x,y,z)$. 
The positions of the 16 iridium (blue and yellow) atoms in the unit cell can be generated from the $Fddd$ space group starting with one iridium atom at $(1/8,1/8,17/24)$ in fractional coordinates.
Iridium atoms are located at the center of edge sharing oxygen (white) octahedral cages.
The three types of nearest neighbor bonds $X$, $Y$, $Z$ are indicated on the left and the altering sign carried by the $\sigma_{\gamma}^{ij}$ symbol, is indicated by $\pm$X and $\pm$Y bonds. The 24-site sublattices
are labeled by a combination of $1,..,6$ and $a,b$.
The spin directions in the counter-rotating spiral at the hidden antiferromagnetic SU(2) point are indicated by red arrows. 
See the main text for details.}
  \label{fig:coorddef}
\end{figure}

The  minimal nearest neighbor $K\Gamma$H model as discussed above is written as\cite{RauPRL2014, LeePRB2015}:
\begin{equation}
H = \sum_{\langle ij\rangle\in \gamma-\text{bond}}[J \mathbf{S}_i \cdot \mathbf{S}_j + K S^\gamma_iS^\gamma_j + \sigma^{ij}_{\gamma} \Gamma(S_i^\alpha S_j^\beta+S^\beta_iS^\alpha_j)],\label{H3D}
\end{equation}
where $i$, $j$ are summing over first nearest neighbours, $\gamma=x,y,z$ denotes the spin component along the $\gamma$-bond and $\alpha,\ \beta $ are the remaining spin components.
Due to the 
rotation of octahedra 
the $\Gamma$ terms on $X$- and $Y$-bond have a sign change indicated by $\sigma^{ij}_{x/y} = \pm 1$, while for the $Z$-bond, $\sigma^{ij}_{z}=+1$ as shown in Fig. \ref{fig:coorddef}.
The space group $Fddd$ describing the hyperhoneycomb lattice allows for a different length of the $Z$-bond from the $X$- and $Y$-bonds, as well as slight distortions of the oxygen octahedra leading to anisotropic strengths between the bond interactions. 
Here we will focus on the ideal case with the same bond interaction strength among X-, Y- and Z-bonds,
and discuss the effects of anisotropy later in the discussion section.

\subsection*{12-site transformation $\mathcal{T}_{12}$ in the hyperhoneycomb}
Previous studies on hidden symmetry via different sets of sublattice transformations on the 2D honeycomb lattice uncovered the zig-zag, stripy, and 120$^\circ$ (vortex) ordered states, in addition to the conventional antiferromagnetic (AF) and ferromagnetic states in the $K\Gamma$H Hamiltonian.\cite{KhaliullinPRB2015}
On the hyperhoneycomb lattice, a 12-site transformation leads to interesting SU(2) symmetric points in the $K\Gamma$ model.
This transformation, dubbed the $\mathcal{T}_{12}$ transformation, is given by,
\begin{gather}
  \mathcal{T}^{a}_6 \\
  \begin{aligned}
    &\text{sublattice $1^{a}$:} \qquad (S^x,S^y,S^z) &\rightarrow & \ \ (\ \ \widetilde{S}^x,\ \ \widetilde{S}^y,\ \ \widetilde{S}^z) \nonumber, \\
    &\text{sublattice $2^{a}$:} \qquad (S^x,S^y,S^z) &\rightarrow & \ \ (\ \ \widetilde{S}^z,-\widetilde{S}^y,\ \ \widetilde{S}^x)\nonumber, \\
    &\text{sublattice $3^{a}$:} \qquad (S^x,S^y,S^z) &\rightarrow & \ \ (-\widetilde{S}^z,-\widetilde{S}^x,\ \ \widetilde{S}^y)\nonumber, \\
    &\text{sublattice $4^{a}$:} \qquad (S^x,S^y,S^z) &\rightarrow & \ \ (\ \ \widetilde{S}^y,\ \ \widetilde{S}^x,-\widetilde{S}^z)\nonumber, \\
    &\text{sublattice $5^{a}$:} \qquad (S^x,S^y,S^z) &\rightarrow & \ \ (-\widetilde{S}^y,\ \ \widetilde{S}^z,-\widetilde{S}^x)\nonumber, \\
    &\text{sublattice $6^{a}$:} \qquad (S^x,S^y,S^z) &\rightarrow & \ \ (-\widetilde{S}^x,-\widetilde{S}^z,-\widetilde{S}^y)\nonumber.
  \end{aligned}
\end{gather}
\begin{gather}
  \mathcal{T}^{b}_6 \\
  \begin{aligned}
    &\text{sublattice $1^{b}$:} \qquad (S^x,S^y,S^z) &\rightarrow & \ \ (-\widetilde{S}^y,-\widetilde{S}^x,-\widetilde{S}^z) \nonumber, \\
    &\text{sublattice $2^{b}$:} \qquad (S^x,S^y,S^z) &\rightarrow & \ \ (\ \ \widetilde{S}^y,-\widetilde{S}^z,-\widetilde{S}^x)\nonumber, \\
    &\text{sublattice $3^{b}$:} \qquad (S^x,S^y,S^z) &\rightarrow & \ \ (\ \ \widetilde{S}^x,\ \ \widetilde{S}^z,-\widetilde{S}^y)\nonumber, \\
    &\text{sublattice $4^{b}$:} \qquad (S^x,S^y,S^z) &\rightarrow & \ \ (-\widetilde{S}^x,-\widetilde{S}^y,\ \ \widetilde{S}^z)\nonumber, \\
    &\text{sublattice $5^{b}$:} \qquad (S^x,S^y,S^z) &\rightarrow & \ \ (-\widetilde{S}^z,\ \ \widetilde{S}^y,\ \ \widetilde{S}^x)\nonumber, \\
    &\text{sublattice $6^{b}$:} \qquad (S^x,S^y,S^z) &\rightarrow & \ \ (\ \ \widetilde{S}^z,\ \ \widetilde{S}^x,\ \ \widetilde{S}^y)\nonumber.
  \end{aligned}
\end{gather}
$\mathcal{T}_{12}$ can be considered as two 6-site transformations denoted by  $\mathcal{T}^{a}_6$ and $\mathcal{T}^{b}_6$.
As shown in Fig. \ref{fig:coorddef}, blue and yellow chains refer to the $a$- and $b$-subset respectively.
The Z-bonds connect the two chains by sites of the same numbering of the $a$- and $b$-subset, i.e.,  $n^a$ and $n^b$ where $n=1,...,6$.
Here $(x,y,z)$ refers to the global axis shown in Fig. \ref{fig:coorddef}.
A pair of two $\mathcal{T}_{12}$ transformations then makes a 24-site sublattice: 
alternating $\mathcal{T}^{a}_6$ and $\mathcal{T}^{b}_6$ along the ${\vec c}$-direction,
i.e., ${\hat x}+{\hat y}$ direction in $(x,y,z)$ coordinates.
For the $K\Gamma$ model, this 12-site transformation reveals a hidden $SU(2)$ symmetry when $K=\Gamma$, where the spin model takes the Heisenberg interaction form in the transformed basis, i.e.,
$H \rightarrow \widetilde{H}= \widetilde{J}\ \sum_{\langle ij\rangle}\widetilde{ \mathbf{S}}_i \cdot \widetilde{ \mathbf{S}}_j$ with ${\widetilde J} = - K$.
For positive $K$ and $\Gamma$, it maps to the FM Heisenberg model, while for negative $K$ and $\Gamma$ it maps to the AF Heisenberg model. 
In the next subsection, we show the corresponding spin ordering for the negative $K$ and $\Gamma$ case, after we transform the AF ordering in the transformed basis back into the original basis. 

\subsection*{counter-rotating spiral order: signature of $SU(2)$ symmetry in 3D $\mathbf{\beta-Li_2IrO_3}$}
As mentioned above, the SU(2) Heisenberg Hamiltonian is found when $K=\Gamma$ in the $\mathcal{T}_{12}$ transformed basis;
$ \widetilde{H}= \widetilde{J}\ \sum_{\langle ij\rangle}\widetilde{ \mathbf{S}}_i \cdot \widetilde{ \mathbf{S}}_j$ with ${\widetilde J} = - K$.
For negative K and $\Gamma$, this leads to AF ordering, which corresponds to the counter-rotating spiral ordering in the original basis as we now show.

\begin{figure}[!ht]
  \centering
  \begin{overpic}[width=0.48\textwidth]{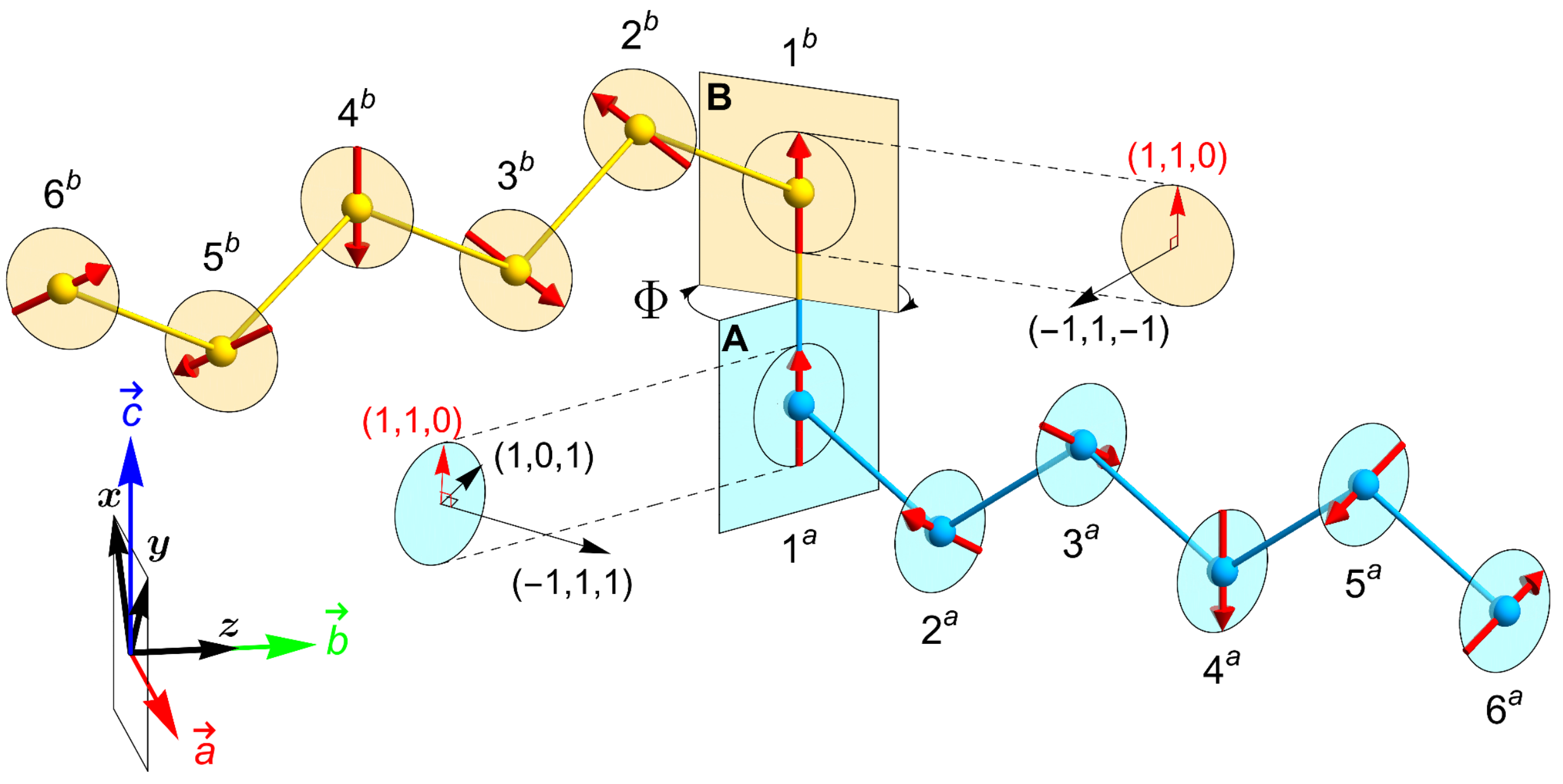}
  \end{overpic}  
  \caption{Pictorial representation of the $\mathcal{T}_{12}$ transformation carried out on all 12-sites. The planes where spin moments rotate are indicated by blue (A) and yellow (B) filled circles. Rotation axes are indicated by black arrows, and the set of red arrows represents an example of moment directions at the exactly solvable SU(2) point.}
  \label{fig:spiralproperties}
\end{figure}

It is straightforward to check that the $\mathcal{T}^a_6$ transformation on the sublattice \{$1^{a}$, $2^{a}$, $3^{a}$ ,$4^{a}$ ,$5^{a}$ ,$6^{a}$\} can be represented by the corresponding set of rotations \{$E$, $C_{2}$, $C^2_{3}$, $C_{3}C_{2}$, $C_{3}$, $C^2_{3}C_{2} $\} where $C_3$ is a $2\pi/3$ rotation about the $(-1,1,1)$-axis and $C_2$ is a $\pi$ rotation about the $(1,0,1)$-axis. These axes are depicted by black arrows on the blue plane in Fig. \ref{fig:spiralproperties}.
Similarly, the $\mathcal{T}^b_6$ transformation on the sublattice \{$1^{b}$, $2^{b}$, $3^{b}$ ,$4^{b}$ ,$5^{b}$ ,$6^{b}$\} can be presented by a different set of rotations \{$C'_{2}$, $C'_{2}C_{2}$, $C'_{2}C^2_{3}$, $C'_{2}C_{3}C_{2}$, $C'_{2}C_{3}$, $C'_{2}C^2_{3}C_{2} $\}  = $C'_2 \times \mathcal{T}^a_6$ where $C'_{2}$ is $\pi$ rotation around the $(-1,1,0)$-axis.

Note that the sublattice \{$1^{a}$, $3^{a}$ ,$5^{a}$\} follows $C_3$ rotations while the sublattice \{$2^{a}$, $4^{a}$, $6^{a}$\} follows $C_3$ rotations in the opposite order, creating a counter-rotating ordering pattern.
Furthermore, the $C_2$ rotation $(1,0,1)$-axis is perpendicular to the $C_3$ rotation $(-1,1,1)$-axis, and thus all 6 moments lie inside one plane named A as shown with a blue plane in Fig. \ref{fig:spiralproperties}.
This generates a counter-rotating spiral order with wavevector $q = 2\pi/3$ along the chain direction made of the subset-$a$.

A similar analysis for $\mathcal{T}^b_6$ with the sublattice \{$1^{b}$, $2^{b}$, $3^{b}$ ,$4^{b}$ ,$5^{b}$ ,$6^{b}$\} transformation could be made.
It is the same as $\mathcal{T}^a_6$ but with an additional $\pi$ rotation around the $(-1,1,0)$-axis denoted by $C'_2$. 
First it contains the opposite orderings of $C_3$ rotations within the 6 sites, like in the $\mathcal{T}^a_6$ transformation, and thus it possesses the counter-rotating pattern.
However, due to the additional $C'_2$ rotation around the $(-1,1,0)$-axis that is applied to all 6 sites, the spiral ordering moments lie in a different plane named B, colored as a yellow plane in Fig. \ref{fig:spiralproperties}.
Note that the blue and yellow planes make an angle of $\Phi=\mathrm{arccos}(1/3) \simeq 70.53^\circ$ as shown in Fig. \ref{fig:spiralproperties}.
This is remarkably close to the angle obtained in neutron scattering analysis.

Given that the transformation maps into an AF Heisenberg model, one then applies a bipartite sign factor $(-1)^n$ on all sites to generate the entire magnetic pattern, where $n = $ even for sites $1^{a/b}$, $3^{a/b}$, $5^{a/b}$, and $n = $ odd for sites  $2^{a/b}$, $4^{a/b}$, $6^{a/b}$. 

Depending on the magnetic ordering moment direction, this generates several different 
counter-rotating spiral orderings.  
The spiral ordering shown by red arrows in Fig. \ref{fig:coorddef} occurs when the moment of $1^{a}$ is pinned along the $(1,1,0)$-direction, which is similar to the experimental observed ordering pattern except for small deviations from the commensurate ordering wave vector.

Below we focus on the parameter space nearby the hidden symmetric point of the $K\Gamma$ model, and perform ED on a 24-site cluster to determine nearby phases.
We also show how the magnetic moment direction is pinned when we go slightly away from the SU(2) point.

\section{Phase diagram and moment direction } \label{phasepinning}

\begin{figure}[!ht]
  \centering
  \begin{overpic}[scale=0.4]{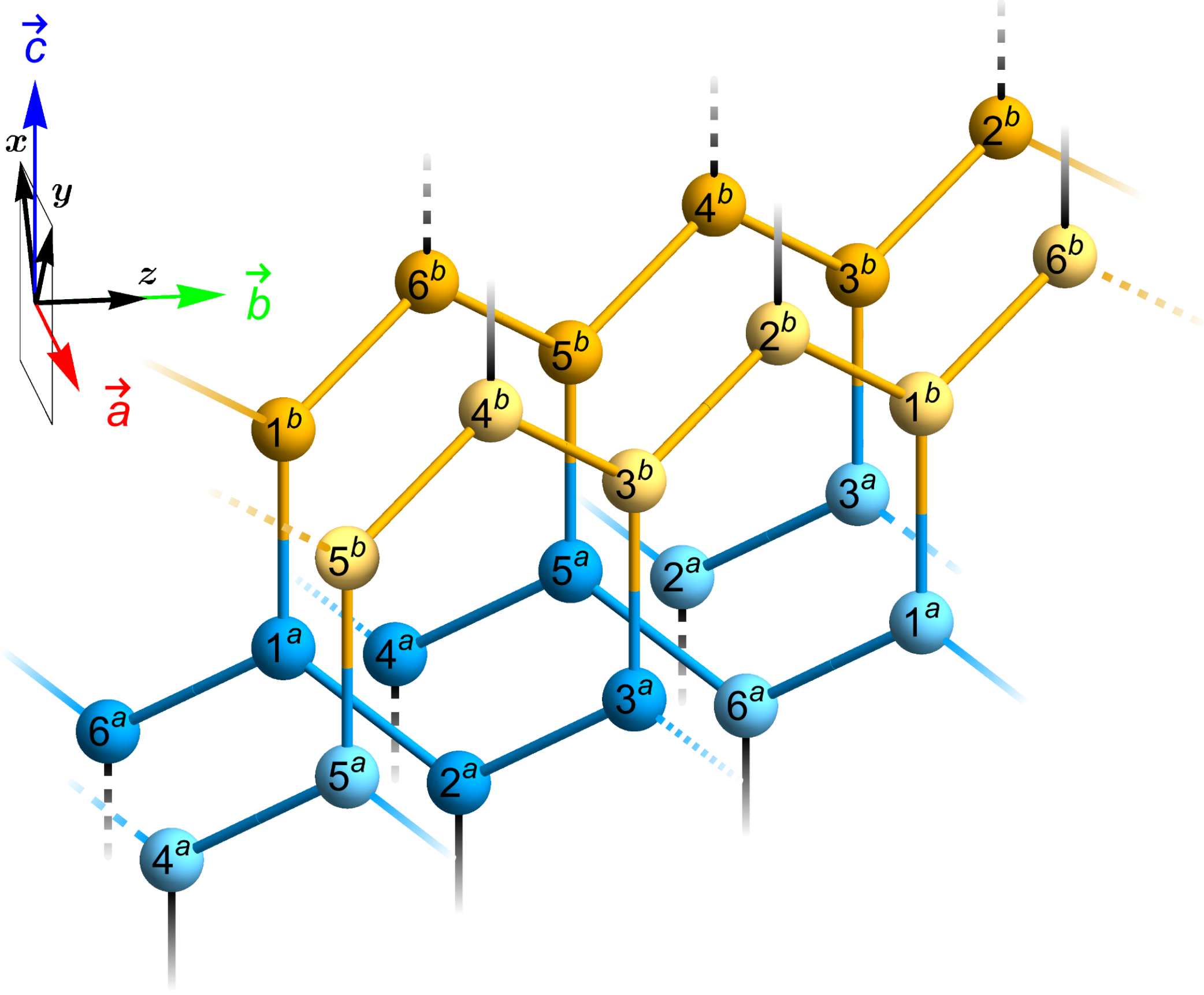}
  \end{overpic}  
  \caption{$N=24$ site cluster used for ED. The blue and yellow colors represent chains where the $\mathcal{T}^a_6$ and $\mathcal{T}^b_6$ are applied respectively. The label $n^{a/b}$ with $n=1,...,6$ indicates the transformation at each site. Periodic boundary condition (PBC) along the chains is
  imposed by connecting the same colour and style of the bonds. For different colour chains, it is imposed by
  connecting sites between $n^a$ and $n^b$ via the $Z$-bond with the same type black lines.
    This boundary condition respects the $\mathcal{T}_{12}$ transformation.
    For the open boundary condition (OBC) discussed in the main text, half the Z-bonds are lost, denoted as black lines along the ${\hat c}$-direction.
    The lighter and darker coloring of both yellow and blue chains indicate the staggering moment convention 
   used in  ${\bf\widetilde{S}}^2_{\text{stg},\mathcal{\overline{T}}_6}$ discussed in Sec. \ref{chains}.
   }
  \label{fig:24T6EDcluster}
\end{figure}

The hidden AF $SU(2)$ point has a central role in understanding the counter-rotating spiral, therefore we explore the parameter space around this point by using ED on the reduced $K\Gamma$ model of Eq. (\ref{H3D}).
Our calculations are performed on the cluster shown in Fig. \ref{fig:24T6EDcluster}, which is a minimal cluster that captures the $\mathcal{T}_{12}$ transformation. 

\subsection*{Phase diagram}
\begin{figure}[!ht]
  \centering
  \begin{overpic}[width=0.48\textwidth]{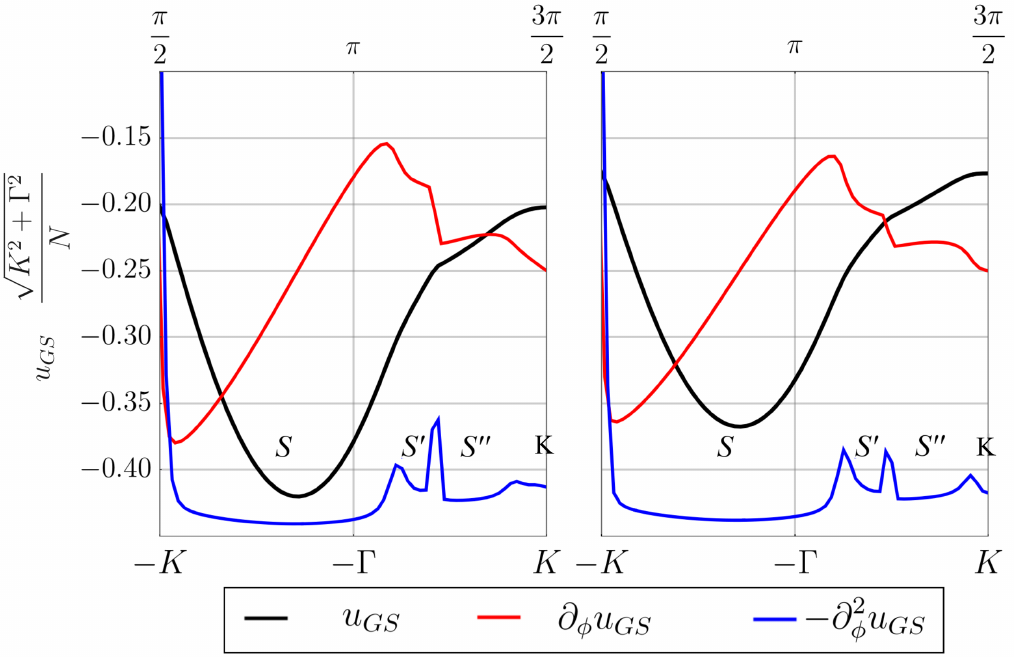}
  \put (27.3,62) {24 site PBC}
  \put (70,62) {24 site OBC}
  \end{overpic}  
  \caption{Phase diagram obtained by ED on the 24-sites cluster shown in Fig. 3 with (a) PBC and (b) OBC. There are four phases identified.
    $K$ refers to the exactly solvable Kitaev spin liquid which is found in the $+K$ region. The $S$ phase is the counter-rotating spiral ordered state, which includes the hidden SU(2) point.
    The $S'$ phase is another magnetically ordered phase and the nature of the $S''$ phase is discussed in the main text.
    $K$ refers to the Kitaev spin liquid phase.}
  \label{fig:24T6EDresults}
\end{figure}

We normalize the exchange strengths by $K=-\sin\phi$ and $\Gamma=\cos\phi$ where $\phi\in[\pi/2,3\pi/2]$.
Phase transitions are identified by singular behavior in the second derivative of the ground state energy density $-\partial^2_{\phi}u_{GS}$, where $u_{GS}=E_{GS}/N$.
Four phases are identified in the region of parameter space following $-K \rightarrow -\Gamma \rightarrow K$, as presented in Fig. \ref{fig:24T6EDresults}.
Our results are insensitive to the choice of boundary condition as can be seen for periodic (PBC) and open boundary conditions (OBC) in Fig. \ref{fig:24T6EDresults}(a) and (b) respectively.

The Kitaev points $-K$ and $K$ exhibit the Kitaev spin liquid state as found in Refs. \citenum{SurendranPRB2009, KimchiPRB2014, LeePRB2014, SchafferPRL2015}. 
%
While the $-K$ point is immediately unstable upon turning on $\Gamma$ interaction, the $+K$ spin liquid, denoted by $K$, occupies a small area of the parameter space. 
%
The counter-rotating spiral phase, denoted by $S$, includes a hidden AF SU(2) point and is extended from the $-K$ limit to slightly beyond the $-\Gamma$ limit. 
Along $-\Gamma \rightarrow K$, two phases $S'$ and $S''$ appear in addition to the Kitaev spin liquid.
The $S'$ phase is magnetically ordered as we discuss in the following section, while the nature of $S''$ is difficult to pin down.
To understand the $S''$ phase, we take the chain limit (1D) and study how the phases evolve as the strength of the $Z$-bond increases
in Sec. IV. Before that, let us understand how the spin moment direction is pinned away from the SU(2) point.

\subsection*{Pinning of magnetic moment direction}
The 12-site transformation naturally gives rise to a counter-rotating spiral, however, the precise ordering pattern depends on the magnetic moment direction on sites $1^{a,b}$.
Neutron scattering data found that the spiral is consistent with a moment pinned along the $(1,1,0)$ direction, as depicted by the red arrows on sites $1^{a,b}$ in Fig. 2.
In this section, we determine the preferred moment direction using ED and discuss the importance of a finite Heisenberg interaction.

To this end, we follow the method used in Ref. \citenum{KhaliullinPRB2016} and construct a spin-coherent product state to form the desired magnetic pattern on the cluster.
This state is then used as an ansatz for the ground state wavefunction.
 In general, for a cluster of $N$ sites labeled by $j$, this ansatz is parameterized by 2N parameters $(\theta_j,\phi_j)$ and takes the form
\begin{equation}
\left| \Psi_{ansatz}(\theta_j,\phi_j) \right> = \prod_{j}{ e^{-i \phi_j S_{j}^z} e^{-i \theta_j S_{j}^y} \left| \uparrow \right>_{j} },
\end{equation}
where there is a spin-$\frac{1}{2}$ coherent state on each site, described by rotations $\theta_j$ and $\phi_j$ about the $y$- and $z$-axes respectively.
It requires a huge parameter space, however, in practice one can reduce the set of 2N parameters if the form of the ground state is known a priori.
For example, an ferromagnetic ordering state needs only two parameters $\theta$ and $\phi$ since all the spins on all sites are aligned in the same direction.
With this ansatz, one can compute the probability $P = |\langle  \Psi_{ansatz} | GS \rangle|^2$ by varying 
$\theta$ and $\phi$, where $|GS\rangle$ is obtained by ED.
The moment direction is then determined by the values $(\theta,\phi)$ which maximize $P$.\cite{KhaliullinPRB2016}


In the same spirit, we can build a wavefunction ansatz at the hidden AF SU(2) point of the $K\Gamma$ model on the hyperhoneycomb when $K = \Gamma$ with negative $K$ and $\Gamma$. 
In the transformed basis, we only need two parameters $(\theta,\phi)$ to describe an AF ansatz.
We then transform back into the original basis via the $\mathcal{T}_{12}$ transformation, thereby generating a counter-rotating spiral ansatz.
Specifically, the AF ansatz in the transformed basis takes the form
\begin{equation}
  | \tilde{\Psi}(\theta,\phi) \rangle = \prod_{j\in \textsf{evens}}{  e^{-i \phi S_{j}^z} e^{-i \theta S_{j}^y} | \tilde{\uparrow}\rangle_{j} e^{-i \phi S_{j+1}^z} e^{-i \theta S_{j+1}^y} | \tilde{\downarrow} \rangle_{j+1}} \label{eq:tildianz},
\end{equation}
where up and down spins are on even and odd sublattices respectively.
Transforming back into the original basis we determine the counter-rotating spiral ansatz,
\begin{align}
U_{\mathcal{T}_{12}} = \prod_{i=0}^{N-1}{ e^{-i \omega_{\mathcal{T}_{12}(i)} \vec{n}_{\mathcal{T}_{12}(i)} \cdot \vec{S}} } \label{eq:t6unit} \\
\left| \Psi_{ansatz}(\theta,\phi) \right> =  U_{\mathcal{T}_{12}}| \tilde{\Psi}(\theta,\phi) \rangle. \label{eq:tildeback}
\end{align}


As expected, we find that all $(\theta,\phi)$ are equally probable at the hidden SU(2) point, indicating no preferred moment direction as shown in Fig. 5 (a).
On the other hand, as one moves away from the SU(2) point, i.e., $|\Gamma| > |K|$, the probability is maximized along the cubic axes, as shown by the representative green arrow in Fig. 5 (b) for the (1,0,0) direction.  
Interestingly, a finite positive Heisenberg term is necessary to pin the direction of the moment along $(1,1,0)$ as shown in Fig. 5 (c).
When a small $J > 0$ is introduced at the SU(2) point, $P(\theta,\phi)$ forms a ring-like shape normal to the $(-1,1,1)$ axis.
This result is similar to a previous study, where it was shown that the classical ground state solution in this parameter regime exhibits an accidental $U(1)$ symmetry.\cite{NataliaPRB2018}
In the present quantum calculation, we find that quantum fluctuations lift this accidental classical degeneracy and selects six equivalent moment directions connected by $C_3$ rotations about the $(-1,1,1)$ axis.
Fig. 5(c) shows that the higher probability regions (in red) associated with these moment directions lie on the ring, with the green arrow pointing in the $(1,1,0)$ direction.

\begin{figure}[!ht]
  \centering
  \begin{overpic}[width=0.45\textwidth]{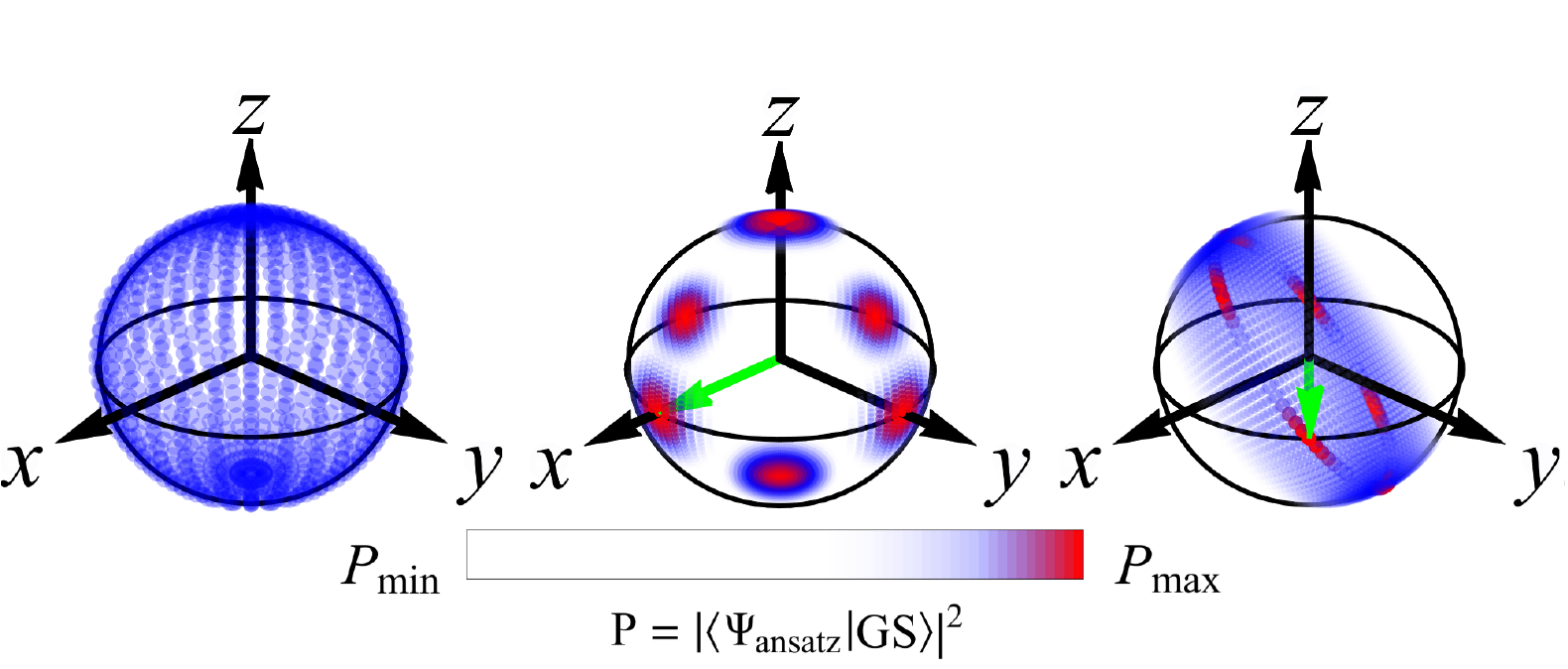}
  \put (13.5,40) {(a)}
  \put (47,40) {(b)}
  \put (81.5,40) {(c)}
  \end{overpic}  
  \caption{Probability maps of $P$. (a) At the hidden $SU(2)$ point $-K=-\Gamma$, no direction is preferred and the probability is equal for every moment direction. (b) Perturbing away from the hidden SU(2) point with $|\Gamma| > |K|$, these bond dependent terms favor the cubic axis for the moment direction, indicated by green arrow for the (1,0,0) direction. (c) Perturbing away from the hidden SU(2) point by $J > 0$, the quantum fluctuations induced by $J$ lifts the accidental classical U(1) degeneracy into six directions related by $C_3$: (1,1,0) (green arrow), (0,1,-1), (-1,0,-1), (-1,-1,0), (0,-1,1), (1,0,1).}
  \label{fig:anztsdensitry}
\end{figure}

\section{Effects of interchain coupling: evolution from 1D to 3D} \label{chains}

As previously stated, in addition to the $S$ phase, which encompasses a large region of phase space in the $K\Gamma$ model, we have found the two Kitaev spin liquid phases at $\pm K$ and have also found two magnetically ordered phases $S'$ and $S''$. 
To gain more insight into the hyperhoneycomb phase diagram, we will investigate the remnants of these phases in the decoupled 1D chain limit, and gradually couple them until the 3D hyperhoneycomb structure is recovered.
To this end, we have performed ED on the cluster in Fig. \ref{fig:24T6EDcluster} with varying strengths of the $Z$-bond that connect the yellow and blue species of 1D chains.
We parameterize the $Z$-bonds (leaving $X$- and $Y$-bonds the same as before) by $K_{z}=-xK = -x\sin\phi$ and $\Gamma_{z}=x\Gamma = x\cos\phi$, where $x \in [0,1]$ is the interchain coupling parameter with $x=0$ and $x=1$ being the 1D and 3D limits, respectively. 

\begin{figure}[!ht]
  \centering
  \begin{overpic}[width=0.5\textwidth]{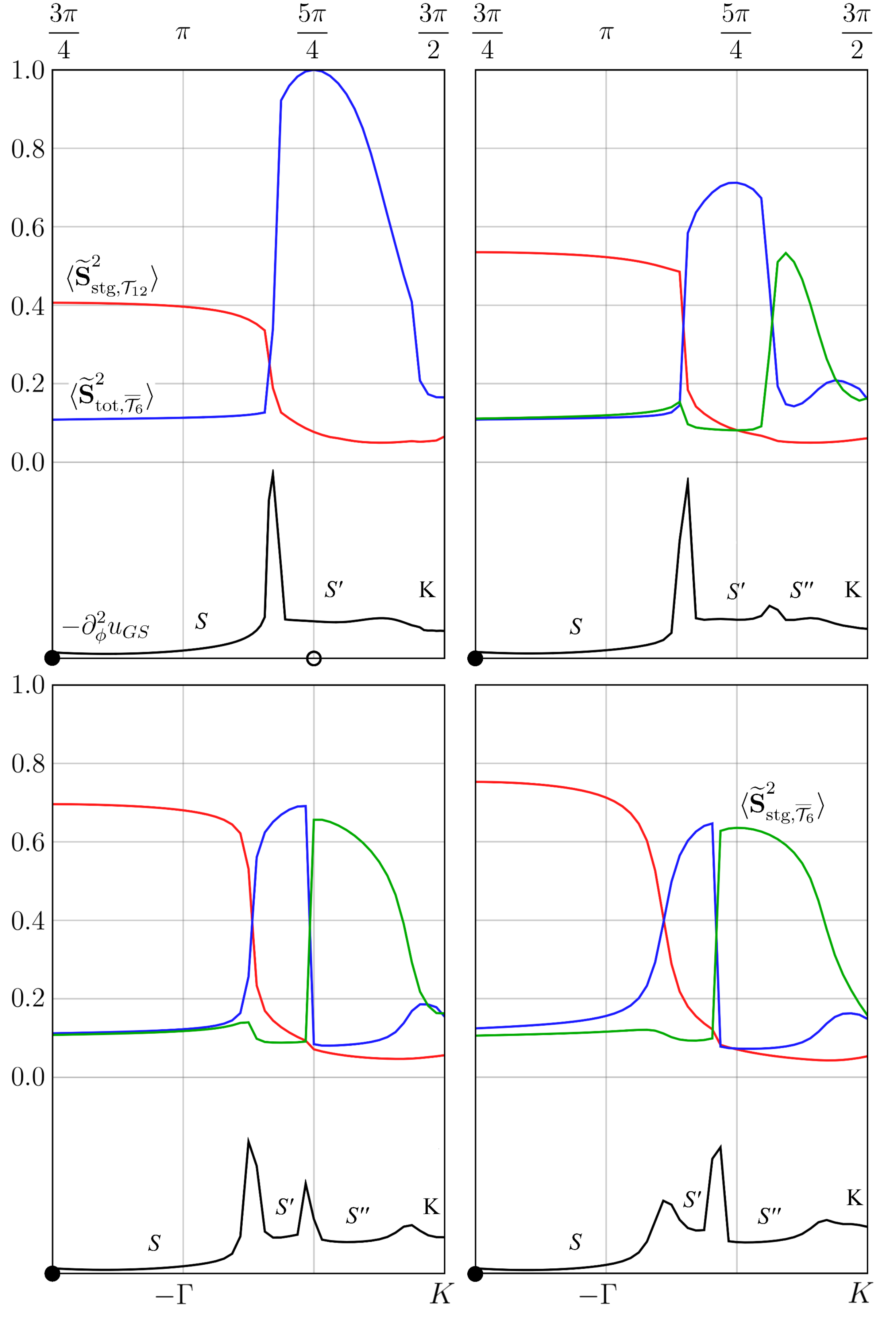}
 \put (4.5,92.1) {\small (a) $x=0.0$}
 \put (36.6,92.1) {\small (b) $x=0.2$}
 \put (4.5,45.5) {\small (c) $x=0.5$}
 \put (36.6,45.5) {\small (d) $x=1.0$}
  \end{overpic}  
  \vspace{-10pt}
  \caption{
  Order parameters $\langle {\bf\widetilde{S}}^2_{\text{stg},\mathcal{T}_{12}}\rangle$ (red), 
 $\langle {\bf\widetilde{S}}^2_{\text{tot},\mathcal{\overline{T}}_6}\rangle$ (blue), and 
 $\langle {\bf\widetilde{S}}^2_{\text{stg},\mathcal{\overline{T}}_6}\rangle$ (green)
 for four values of interchain coupling $x$ are shown. 
 The dark solid line is the second derivative of ground state energy with respect to $\phi$ to represent the phase transitions.
 The hidden AFM SU(2) point from $\mathcal{T}_{12}$ transformation is shown as a filled circle in (a)-(d), and the hidden FM SU(2) point
 from $\mathcal{\overline{T}}_{6}$  as an empty circle in the 1D limit where $x=0$. 
 }
 \label{fig:1dto3d}
\end{figure}
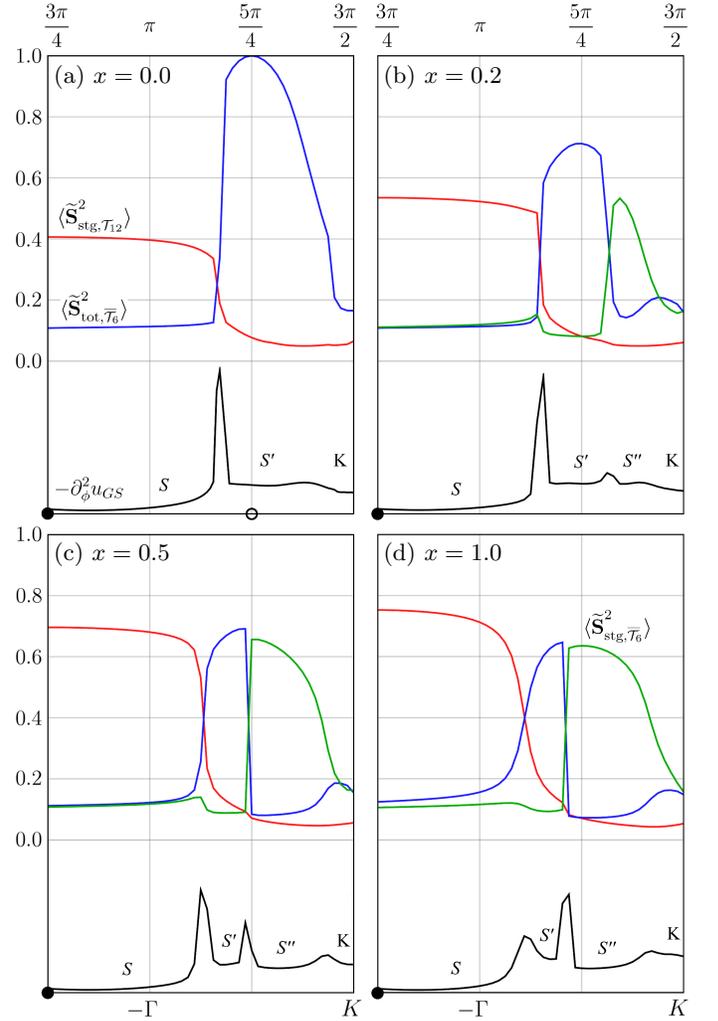

The $x=0$ limit was studied with ED on chains of up to $N=30$ sites and phase boundaries were determined by identifying singularities in $-\partial^2_\phi u_{GS}$.
Our results are shown in Fig. \ref{fig:1dto3d}(a) when $\Gamma < 0$.
%
%
Remarkably, the decoupled chain phase diagram when $\Gamma < 0$ is very similar to the 3D limit.
We find four phases in total, two near the $\pm K$ limits and the 1D analogue of the $S$ and $S'$ phases, on which we elaborate below.

The $\pm K$ limits, corresponding to the 1D Kitaev spin chain, have been studied previously in Ref. \citenum{XiangPRL2007}.
In these limits, the model can be solved exactly by Jordan-Wigner transformation, and the resulting quasiparticle excitations are Majorana fermions with a gapless dispersion.
Here we find that the ferromagnetic Kitaev limit is immediately unstable upon perturbation by any finite $\Gamma$; while by contrast, the antiferromagnetic Kitaev limit is stable for sufficiently small $\Gamma$.
This behavior in 1D is in agreement with our results in the 3D limit near the Kitaev limits. 

To understand the $S$ and $S'$ phases in the 1D limit, we make use of two different types of sublattice transformations which reveal four points of hidden SU(2) symmetry in 1D.
As alluded to previously, the 6-site transformations $\mathcal{T}_6^a$ and $\mathcal{T}_6^b$ defined in section \ref{setup_and_su2_points} can be applied to each decoupled chain to yield two points of hidden SU(2) symmetry when $\pm K = \pm\Gamma$.
For the negative case of these is in the 1D $S$ phase and corresponds to the AF Heisenberg limit, where the ground state is quasi-long-range-ordered owing to the low dimensionality.
Special to 1D, we may further define an additional set of 6-site transformations $\mathcal{{\overline{T}}}_6^{a,b}$ by,
\begin{gather}
  \mathcal{{\overline{T}}}^{a}_6 \\
  \begin{aligned}
    &\text{sublattice $1^{a}$:} \qquad (S^x,S^y,S^z) &\rightarrow & \ \ (\ \ \widetilde{S}^x,\ \ \widetilde{S}^y,\ \ \widetilde{S}^z) \nonumber, \\
    &\text{sublattice $2^{a}$:} \qquad (S^x,S^y,S^z) &\rightarrow & \ \ (-\widetilde{S}^z,-\widetilde{S}^y,-\widetilde{S}^x)\nonumber, \\
    &\text{sublattice $3^{a}$:} \qquad (S^x,S^y,S^z) &\rightarrow & \ \ (\ \ \widetilde{S}^z,-\widetilde{S}^x,-\widetilde{S}^y)\nonumber, \\
    &\text{sublattice $4^{a}$:} \qquad (S^x,S^y,S^z) &\rightarrow & \ \ (\ \ \widetilde{S}^y,\ \ \widetilde{S}^x,-\widetilde{S}^z)\nonumber, \\
    &\text{sublattice $5^{a}$:} \qquad (S^x,S^y,S^z) &\rightarrow & \ \ (-\widetilde{S}^y,-\widetilde{S}^z,\ \ \widetilde{S}^x)\nonumber, \\
    &\text{sublattice $6^{a}$:} \qquad (S^x,S^y,S^z) &\rightarrow & \ \ (-\widetilde{S}^x,\ \ \widetilde{S}^z,\ \ \widetilde{S}^y)\nonumber.
  \end{aligned}
\end{gather}
\begin{gather}
  \mathcal{{\overline{T}}}^{b}_6 \\
  \begin{aligned}
    &\text{sublattice $1^{b}$:} \qquad (S^x,S^y,S^z) &\rightarrow & \ \ (\ \ \widetilde{S}^y,\ \ \widetilde{S}^x,-\widetilde{S}^z) \nonumber, \\
    &\text{sublattice $2^{b}$:} \qquad (S^x,S^y,S^z) &\rightarrow & \ \ (-\widetilde{S}^y,\ \ \widetilde{S}^z,-\widetilde{S}^x)\nonumber, \\
    &\text{sublattice $3^{b}$:} \qquad (S^x,S^y,S^z) &\rightarrow & \ \ (-\widetilde{S}^x,-\widetilde{S}^z,-\widetilde{S}^y)\nonumber, \\
    &\text{sublattice $4^{b}$:} \qquad (S^x,S^y,S^z) &\rightarrow & \ \ (\ \ \widetilde{S}^x,\ \ \widetilde{S}^y,\ \ \widetilde{S}^z)\nonumber, \\
    &\text{sublattice $5^{b}$:} \qquad (S^x,S^y,S^z) &\rightarrow & \ \ (\ \ \widetilde{S}^z,-\widetilde{S}^y,\ \ \widetilde{S}^x)\nonumber, \\
    &\text{sublattice $6^{b}$:} \qquad (S^x,S^y,S^z) &\rightarrow & \ \ (-\widetilde{S}^z,-\widetilde{S}^x,\ \ \widetilde{S}^y)\nonumber.
  \end{aligned}
\end{gather}

As a consequence of these transformations, the $K\Gamma$ model on the chain is mapped to $\bar{H} = \sum_{\langle i,j \rangle} J{\bf \widetilde{S}}_i\cdot {\bf \widetilde{S}}_j$ with $J = \Gamma$ when $\pm\Gamma=\mp K$.
For the same reasons as with $\mathcal{T}_6^{a,b}$, the different subsets of a- and b-type chains are due to the sign convention of $X$ and $Y$ bonds as shown in Fig. \ref{fig:coorddef}, and one must apply the appropriate $\mathcal{\overline{T}}_6$ for each chain.
Thus we have found that the $S'$ phase in 1D is ordered due to the point of hidden FM SU(2) symmetry when $-\Gamma = K$, with $K > 0$.
It has no net magnetization, but possesses a finite sublattice magnetization.

To understand how the $S$ and $S'$ phases evolve from 1D to 3D, we compute the expectation values of
the squared total spin in the transformed basis for each of the phases $S$ and $S'$, 
denoted by $\langle {\bf\widetilde{S}}^2_{\text{stg},\mathcal{T}_{12}}\rangle$ and
 $\langle {\bf\widetilde{S}}^2_{\text{tot},\mathcal{\overline{T}}_6}\rangle$ respectively.
In the calculation of squared total spin for the case of the $S$ phase, we further take care to introduce a relative staggered sign $(-1)^i$
between the two sublattices of the 1D chain in order to capture the AF nature of the phase owing to the hidden AF SU(2) point.
We plot $\langle {\bf\widetilde{S}}^2_{\text{stg},\mathcal{T}_{12}}\rangle $ (red line) and 
$\langle {\bf\widetilde{S}}^2_{\text{tot},\mathcal{\overline{T}}_6}\rangle$ (blue line) in Fig. \ref{fig:1dto3d}(a) and have normalized to the maximum total value expected for a FM, i.e. $\frac{N}{2}(\frac{N}{2}+1)$.
We find that $S$ and $S'$ are well differentiated by these order parameters and that the $S'$ phase saturates the order parameter when $K=-\Gamma$, as expected due to the hidden FM nature of the phase.
Accordingly, the $S$ phase is also well described, however the staggered squared total spin does not saturate the order parameter because of the strong AF quantum fluctuations in 1D. 

When the interchain coupling is turned on, i.e., $x \neq 0$, the $\mathcal{ T}_{12}$ transformations
maps  the $K\Gamma$ model to the AF Heisenberg model when $K=\Gamma$ with K and $ \Gamma < 0$, 
as denoted by a black dot in Fig. \ref{fig:1dto3d}.
As a result the $S$ phase is present from the 1D limit with $x=0$ up to the 3D limit with $x=1$. 
On the other hand, when $x \neq 0$ the $\mathcal{\overline{T}}_6$ transformations no longer maps
to the Heisenberg model because the transformations do not satisfy the $Z$-bonds except on sites $1^{a,b}$ and $4^{a,b}$.
Despite that, the $S'$ phase survives for all $x \neq 0$, as shown in Fig. \ref{fig:1dto3d}(b)-(d), even though the phase space becomes narrower.
Due to the frustration of $Z$-bonds, a new phase $S''$, between the $S'$ and the AF Kitaev spin liquid phases appears. 

In the $S''$ phase, since the a- and b-subset chains independently possess the hidden FM SU(2) symmetry,
the $X$- and $Y$-bond interactions try to maximize FM ordering in the transformed basis, when the $Z$-bond interaction is small, i.e., $x \ll 1$. 
Thus we investigate the tendency of FM ordering in $\mathcal{\overline{T}}_6$ basis along the chains.
To avoid the frustration of $Z$-bond, it is likely that spins on different chains may orient in opposite directions such as
AF ordering between chains. The AF ordering between chains is denoted by the lighter and darker colouring of chains in Fig. 3.
We compute the corresponding staggered squared total spin $\langle {\bf\widetilde{S}}^2_{\text{stg},\mathcal{\overline{T}}_6}\rangle $.
The green line referring to the order parameter is plotted in Fig. \ref{fig:1dto3d}(b)-(d), which captures the nature of this ordered phase. 

\section{Summary and Discussion}

The distinct properties of the Kitaev spin liquid, an example of topological order in which the electron's spin-$\frac{1}{2}$ fractionalizes into Majorana and $\mathbb{Z}_2$ flux excitations, sets it apart from the more common magnetically ordered states found in real materials.
Indeed, the exotic nature of the Kitaev spin liquid has motivated the community to attempt to realize the phase in numerous candidate materials from the 2D honeycomb materials A$_2$IrO$_3$ (A = Li, Ni) and $\alpha$-RuCl$_3$ to the 3D hyperhoneycomb materials $\beta$,$\gamma$-LiIrO$_3$.
Despite this intensive search, all these materials magnetically order and, in particular, an intriguing counter-rotating spiral is found in the Li-based iridates.
It is thus important in these materials to understand the role of the competing interactions in the minimal nearest neighbor K$\Gamma$H with potentially further neighbor interactions.

In this work, we have focused on the counter-rotating noncoplanar spiral order in 3D Iridates.
Previous studies pertaining to $\beta$-Li$_2$IrO$_3$ have analyzed the classical K$\Gamma$H spin model and determined, based on the symmetry of the classical ground state, that the experimentally observed magnetic order could be in a region of parameter space in which $-\Gamma \sim -K$, with $J>0$.
However, there is still little insight into the preferred local moment direction in $\beta$-Li$_2$IrO$_3$ and moroever, these studies do not give a fully quantum mechanical treatment of the model owing to its inherent difficulty in a 3D system such as this.
Here we present a 12-site transformation gives an exactly solvable point of hidden AF SU(2) symmetry when $-K = -\Gamma$.
Using this as our guide, we have been able to construct the minimal cluster necessary to use in ED to understand the counter-rotating spiral phase as well as nearby phases.

In addition to the Kitaev spin liquid phases, three magnetically ordered phases $S$, $S'$ and $S''$ are found.
Apart from the $S''$ phase,  
each of these phases has a 1D analogue if we tune the hyperhoneycomb structure to the limit of decoupled chains.
The $S$ phase is a counter-rotating spiral ordered state, which is a direct consequence of the $\mathcal{T}_{12}$ sublattice transformation.
Furthermore at $-K = -\Gamma$, a small $J > 0$ is added and
the local moment is pinned to in the (1,1,0) direction or its symmetry equivalents.
These results are consistent with the neutron scattering experiments on $\beta$-Li$_2$IrO$_3$.
The nearby $S'$ and $S''$ phases are also interesting.
We see that the $S'$ phase is a spiral whose pattern is determined by another set of 6-site transformations $\mathcal{\overline{T}}_6^{a,b}$, which give rise to two new points of hidden SU(2) symmetry when $\pm\Gamma = \mp K$ in the decoupled chain limit.
However the mappings of Heisenberg model by $\mathcal{\overline{T}}_6^{a,b}$ transformations work only for the chains separately, 
and the $Z$-bonds are frustrated when the chains are coupled
As a result of  $Z$-bond frustration, another ordered phase $S''$, distinct from $S$ and $S'$ emerges between $S'$ and Kitaev spin liquid phase. 
However, the phase transition between $S''$ and the spin liquid phase is rather unclear which may be related to a finite size of the system
and choice of clusters. Note that the 24-site ED with OBC in Fig. 4 (b) shows a clearer signature of phase transition than that of PBC in (a).
It will be interesting to study the effects of anisotropy by cutting $Z$-bonds in both $S''$ and the KSL phases in the 3D system, which may 
change the transition between the two, similar to a recent study from the positive $\Gamma$ to a negative Kitaev 
point in 2D honeycomb lattice.\cite{AndreiNPJ2018}

The mapping to the Heisenberg model by $\mathcal{T}_{12}$ sublattice transformation is exact
at $\pm K= \pm \Gamma$ without $J$. The corresponding ordered $S$ phase indeed captures the spiral ordering,
but  how the $S$ phase becomes incommensurate in the presence of $J$ is not clear,
due to the finite size of the cluster although this problem plagues all finite size ED calculations.
On the other hand, previous classical studies have shown that the $S$ spiral phase, when considering only negative $K$ and $\Gamma$ terms, has a constant ordering vector of $(2/3, 0,0)$ and a finite $J$ is enough to induce incommensuration. \cite{LeePRB2014} Second nearest neighbor Heisemberg $J_2$ term has also been found to cause incommensuration. \cite{VishwanathPRB2015} 
A quantum investigation of these phenomenon would be an also interesting topic for future study.


Our study suggests that $\beta$-Li$_2$IrO$_3$ is likely within the magnetically ordered $S$ phase.
To push the system into a spin liquid state, one would have to shift towards the relatively stable $+K$ point and applying hydrostatic pressure could be a way to achieve this.
Pressure measurements on the 3D iridates indicate that the magnetic ordering disappears in the intermediate pressure range,
while the structure transition occurs at a higher pressure.\cite{VeigaPRB2017,BreznayPRB2017,GegenwartPRL2018} This implies a new ground state is achieved by applying hydrostatic pressures,
and further experimental and theoretical studies to understand the intermediate phase are required. 
Interestingly, \textit{ab initio} pressure studies have shown that the Kitaev term can become positive while $\Gamma$ remains negative.\cite{KimPRB2016}
However, the interaction anisotropy is also enhanced under pressure, and thus one needs to
investigate its effects on both ordered and spin liquid states as discussed above.
Uniaxial stress may be another way to effectively tune the microscopic interactions in the laboratory setting,
  which remains a challenging task.


\begin{acknowledgements}
This work was supported by the Natural Sciences and Engineering Research Council of Canada and the Center for Quantum Materials at the University of Toronto. Computations were performed on the GPC and Niagara supercomputers at the SciNet HPC Consortium. SciNet is funded by: the Canada Foundation for Innovation under the auspices of Compute Canada; the Government of Ontario; Ontario Research Fund - Research Excellence; and the University of Toronto.
\end{acknowledgements}

\newpage
\bibliography{KGmodel}

\begin{thebibliography}{45}%
\makeatletter
\providecommand \@ifxundefined [1]{%
 \@ifx{#1\undefined}
}%
\providecommand \@ifnum [1]{%
 \ifnum #1\expandafter \@firstoftwo
 \else \expandafter \@secondoftwo
 \fi
}%
\providecommand \@ifx [1]{%
 \ifx #1\expandafter \@firstoftwo
 \else \expandafter \@secondoftwo
 \fi
}%
\providecommand \natexlab [1]{#1}%
\providecommand \enquote  [1]{``#1''}%
\providecommand \bibnamefont  [1]{#1}%
\providecommand \bibfnamefont [1]{#1}%
\providecommand \citenamefont [1]{#1}%
\providecommand \href@noop [0]{\@secondoftwo}%
\providecommand \href [0]{\begingroup \@sanitize@url \@href}%
\providecommand \@href[1]{\@@startlink{#1}\@@href}%
\providecommand \@@href[1]{\endgroup#1\@@endlink}%
\providecommand \@sanitize@url [0]{\catcode `\\12\catcode `\$12\catcode
  `\&12\catcode `\#12\catcode `\^12\catcode `\_12\catcode `\%12\relax}%
\providecommand \@@startlink[1]{}%
\providecommand \@@endlink[0]{}%
\providecommand \url  [0]{\begingroup\@sanitize@url \@url }%
\providecommand \@url [1]{\endgroup\@href {#1}{\urlprefix }}%
\providecommand \urlprefix  [0]{URL }%
\providecommand \Eprint [0]{\href }%
\providecommand \doibase [0]{http://dx.doi.org/}%
\providecommand \selectlanguage [0]{\@gobble}%
\providecommand \bibinfo  [0]{\@secondoftwo}%
\providecommand \bibfield  [0]{\@secondoftwo}%
\providecommand \translation [1]{[#1]}%
\providecommand \BibitemOpen [0]{}%
\providecommand \bibitemStop [0]{}%
\providecommand \bibitemNoStop [0]{.\EOS\space}%
\providecommand \EOS [0]{\spacefactor3000\relax}%
\providecommand \BibitemShut  [1]{\csname bibitem#1\endcsname}%
\let\auto@bib@innerbib\@empty
\bibitem [{\citenamefont {Anderson}(1973)}]{AndersonMRB1973}%
  \BibitemOpen
  \bibfield  {author} {\bibinfo {author} {\bibfnamefont {P.}~\bibnamefont
  {Anderson}},\ }\href
  {http://www.sciencedirect.com/science/article/pii/0025540873901670}
  {\bibfield  {journal} {\bibinfo  {journal} {Materials Research Bulletin}\
  }\textbf {\bibinfo {volume} {8}},\ \bibinfo {pages} {153 } (\bibinfo {year}
  {1973})}\BibitemShut {NoStop}%
\bibitem [{\citenamefont {Balents}(2010)}]{BalentsNat2010}%
  \BibitemOpen
  \bibfield  {author} {\bibinfo {author} {\bibfnamefont {L.}~\bibnamefont
  {Balents}},\ }\href {http://dx.doi.org/10.1038/nature08917} {\bibfield
  {journal} {\bibinfo  {journal} {Nature}\ }\textbf {\bibinfo {volume} {464,
  199}} (\bibinfo {year} {2010})}\BibitemShut {NoStop}%
\bibitem [{\citenamefont {Kitaev}(2006)}]{KitaevAP2006}%
  \BibitemOpen
  \bibfield  {author} {\bibinfo {author} {\bibfnamefont {A.}~\bibnamefont
  {Kitaev}},\ }\href
  {http://www.sciencedirect.com/science/article/pii/S0003491605002381"}
  {\bibfield  {journal} {\bibinfo  {journal} {Annals of Physics}\ }\textbf
  {\bibinfo {volume} {321}},\ \bibinfo {pages} {2 } (\bibinfo {year}
  {2006})}\BibitemShut {NoStop}%
\bibitem [{\citenamefont {Jackeli}\ and\ \citenamefont
  {Khaliullin}(2009)}]{JackeliPRL2009}%
  \BibitemOpen
  \bibfield  {author} {\bibinfo {author} {\bibfnamefont {G.}~\bibnamefont
  {Jackeli}}\ and\ \bibinfo {author} {\bibfnamefont {G.}~\bibnamefont
  {Khaliullin}},\ }\href {\doibase 10.1103/PhysRevLett.102.017205} {\bibfield
  {journal} {\bibinfo  {journal} {Phys. Rev. Lett.}\ }\textbf {\bibinfo
  {volume} {102}},\ \bibinfo {pages} {017205} (\bibinfo {year}
  {2009})}\BibitemShut {NoStop}%
\bibitem [{\citenamefont {Singh}\ and\ \citenamefont
  {Gegenwart}(2010)}]{SinghPRB2010}%
  \BibitemOpen
  \bibfield  {author} {\bibinfo {author} {\bibfnamefont {Y.}~\bibnamefont
  {Singh}}\ and\ \bibinfo {author} {\bibfnamefont {P.}~\bibnamefont
  {Gegenwart}},\ }\href {\doibase 10.1103/PhysRevB.82.064412} {\bibfield
  {journal} {\bibinfo  {journal} {Phys. Rev. B}\ }\textbf {\bibinfo {volume}
  {82}},\ \bibinfo {pages} {064412} (\bibinfo {year} {2010})}\BibitemShut
  {NoStop}%
\bibitem [{\citenamefont {Singh}\ \emph {et~al.}(2012)\citenamefont {Singh},
  \citenamefont {Manni}, \citenamefont {Reuther}, \citenamefont {Berlijn},
  \citenamefont {Thomale}, \citenamefont {Ku}, \citenamefont {Trebst},\ and\
  \citenamefont {Gegenwart}}]{GegenwartPRL2012}%
  \BibitemOpen
  \bibfield  {author} {\bibinfo {author} {\bibfnamefont {Y.}~\bibnamefont
  {Singh}}, \bibinfo {author} {\bibfnamefont {S.}~\bibnamefont {Manni}},
  \bibinfo {author} {\bibfnamefont {J.}~\bibnamefont {Reuther}}, \bibinfo
  {author} {\bibfnamefont {T.}~\bibnamefont {Berlijn}}, \bibinfo {author}
  {\bibfnamefont {R.}~\bibnamefont {Thomale}}, \bibinfo {author} {\bibfnamefont
  {W.}~\bibnamefont {Ku}}, \bibinfo {author} {\bibfnamefont {S.}~\bibnamefont
  {Trebst}}, \ and\ \bibinfo {author} {\bibfnamefont {P.}~\bibnamefont
  {Gegenwart}},\ }\href {\doibase 10.1103/PhysRevLett.108.127203} {\bibfield
  {journal} {\bibinfo  {journal} {Phys. Rev. Lett.}\ }\textbf {\bibinfo
  {volume} {108}},\ \bibinfo {pages} {127203} (\bibinfo {year}
  {2012})}\BibitemShut {NoStop}%
\bibitem [{\citenamefont {Plumb}\ \emph {et~al.}(2014)\citenamefont {Plumb},
  \citenamefont {Clancy}, \citenamefont {Sandilands}, \citenamefont {Shankar},
  \citenamefont {Hu}, \citenamefont {Burch}, \citenamefont {Kee},\ and\
  \citenamefont {Kim}}]{KimPRB2014}%
  \BibitemOpen
  \bibfield  {author} {\bibinfo {author} {\bibfnamefont {K.~W.}\ \bibnamefont
  {Plumb}}, \bibinfo {author} {\bibfnamefont {J.~P.}\ \bibnamefont {Clancy}},
  \bibinfo {author} {\bibfnamefont {L.~J.}\ \bibnamefont {Sandilands}},
  \bibinfo {author} {\bibfnamefont {V.~V.}\ \bibnamefont {Shankar}}, \bibinfo
  {author} {\bibfnamefont {Y.~F.}\ \bibnamefont {Hu}}, \bibinfo {author}
  {\bibfnamefont {K.~S.}\ \bibnamefont {Burch}}, \bibinfo {author}
  {\bibfnamefont {H.-Y.}\ \bibnamefont {Kee}}, \ and\ \bibinfo {author}
  {\bibfnamefont {Y.-J.}\ \bibnamefont {Kim}},\ }\href {\doibase
  10.1103/PhysRevB.90.041112} {\bibfield  {journal} {\bibinfo  {journal} {Phys.
  Rev. B}\ }\textbf {\bibinfo {volume} {90}},\ \bibinfo {pages} {041112}
  (\bibinfo {year} {2014})}\BibitemShut {NoStop}%
\bibitem [{\citenamefont {Kim}\ \emph {et~al.}(2015{\natexlab{a}})\citenamefont
  {Kim}, \citenamefont {V.}, \citenamefont {Catuneanu},\ and\ \citenamefont
  {Kee}}]{KeePRB2015}%
  \BibitemOpen
  \bibfield  {author} {\bibinfo {author} {\bibfnamefont {H.-S.}\ \bibnamefont
  {Kim}}, \bibinfo {author} {\bibfnamefont {V.~S.}\ \bibnamefont {V.}},
  \bibinfo {author} {\bibfnamefont {A.}~\bibnamefont {Catuneanu}}, \ and\
  \bibinfo {author} {\bibfnamefont {H.-Y.}\ \bibnamefont {Kee}},\ }\href
  {\doibase 10.1103/PhysRevB.91.241110} {\bibfield  {journal} {\bibinfo
  {journal} {Phys. Rev. B}\ }\textbf {\bibinfo {volume} {91}},\ \bibinfo
  {pages} {241110} (\bibinfo {year} {2015}{\natexlab{a}})}\BibitemShut
  {NoStop}%
\bibitem [{\citenamefont {Modic}\ \emph {et~al.}(2014)\citenamefont {Modic},
  \citenamefont {Smidt}, \citenamefont {Kimchi}, \citenamefont {Breznay},
  \citenamefont {Biffin}, \citenamefont {Choi}, \citenamefont {Johnson},
  \citenamefont {Coldea}, \citenamefont {Watkins-Curry}, \citenamefont
  {McCandless}, \citenamefont {Chan}, \citenamefont {Gandara}, \citenamefont
  {Islam}, \citenamefont {Vishwanath}, \citenamefont {Shekhter}, \citenamefont
  {McDonald},\ and\ \citenamefont {Analytis}}]{AnalytisNC2014}%
  \BibitemOpen
  \bibfield  {author} {\bibinfo {author} {\bibfnamefont {K.~A.}\ \bibnamefont
  {Modic}}, \bibinfo {author} {\bibfnamefont {T.~E.}\ \bibnamefont {Smidt}},
  \bibinfo {author} {\bibfnamefont {I.}~\bibnamefont {Kimchi}}, \bibinfo
  {author} {\bibfnamefont {N.~P.}\ \bibnamefont {Breznay}}, \bibinfo {author}
  {\bibfnamefont {A.}~\bibnamefont {Biffin}}, \bibinfo {author} {\bibfnamefont
  {S.}~\bibnamefont {Choi}}, \bibinfo {author} {\bibfnamefont {R.~D.}\
  \bibnamefont {Johnson}}, \bibinfo {author} {\bibfnamefont {R.}~\bibnamefont
  {Coldea}}, \bibinfo {author} {\bibfnamefont {P.}~\bibnamefont
  {Watkins-Curry}}, \bibinfo {author} {\bibfnamefont {G.~T.}\ \bibnamefont
  {McCandless}}, \bibinfo {author} {\bibfnamefont {J.~Y.}\ \bibnamefont
  {Chan}}, \bibinfo {author} {\bibfnamefont {F.}~\bibnamefont {Gandara}},
  \bibinfo {author} {\bibfnamefont {Z.}~\bibnamefont {Islam}}, \bibinfo
  {author} {\bibfnamefont {A.}~\bibnamefont {Vishwanath}}, \bibinfo {author}
  {\bibfnamefont {A.}~\bibnamefont {Shekhter}}, \bibinfo {author}
  {\bibfnamefont {R.~D.}\ \bibnamefont {McDonald}}, \ and\ \bibinfo {author}
  {\bibfnamefont {J.~G.}\ \bibnamefont {Analytis}},\ }\href {\doibase
  10.1038/ncomms5203} {\bibfield  {journal} {\bibinfo  {journal} {Nature
  Communications}\ }\textbf {\bibinfo {volume} {5}},\ \bibinfo {pages} {4203}
  (\bibinfo {year} {2014})}\BibitemShut {NoStop}%
\bibitem [{\citenamefont {Takayama}\ \emph {et~al.}(2015)\citenamefont
  {Takayama}, \citenamefont {Kato}, \citenamefont {Dinnebier}, \citenamefont
  {Nuss}, \citenamefont {Kono}, \citenamefont {Veiga}, \citenamefont {Fabbris},
  \citenamefont {Haskel},\ and\ \citenamefont {Takagi}}]{TakagiPRL2015}%
  \BibitemOpen
  \bibfield  {author} {\bibinfo {author} {\bibfnamefont {T.}~\bibnamefont
  {Takayama}}, \bibinfo {author} {\bibfnamefont {A.}~\bibnamefont {Kato}},
  \bibinfo {author} {\bibfnamefont {R.}~\bibnamefont {Dinnebier}}, \bibinfo
  {author} {\bibfnamefont {J.}~\bibnamefont {Nuss}}, \bibinfo {author}
  {\bibfnamefont {H.}~\bibnamefont {Kono}}, \bibinfo {author} {\bibfnamefont
  {L.~S.~I.}\ \bibnamefont {Veiga}}, \bibinfo {author} {\bibfnamefont
  {G.}~\bibnamefont {Fabbris}}, \bibinfo {author} {\bibfnamefont
  {D.}~\bibnamefont {Haskel}}, \ and\ \bibinfo {author} {\bibfnamefont
  {H.}~\bibnamefont {Takagi}},\ }\href {\doibase
  10.1103/PhysRevLett.114.077202} {\bibfield  {journal} {\bibinfo  {journal}
  {Phys. Rev. Lett.}\ }\textbf {\bibinfo {volume} {114}},\ \bibinfo {pages}
  {077202} (\bibinfo {year} {2015})}\BibitemShut {NoStop}%
\bibitem [{\citenamefont {Liu}\ \emph {et~al.}(2011)\citenamefont {Liu},
  \citenamefont {Berlijn}, \citenamefont {Yin}, \citenamefont {Ku},
  \citenamefont {Tsvelik}, \citenamefont {Kim}, \citenamefont {Gretarsson},
  \citenamefont {Singh}, \citenamefont {Gegenwart},\ and\ \citenamefont
  {Hill}}]{HillPRB2011}%
  \BibitemOpen
  \bibfield  {author} {\bibinfo {author} {\bibfnamefont {X.}~\bibnamefont
  {Liu}}, \bibinfo {author} {\bibfnamefont {T.}~\bibnamefont {Berlijn}},
  \bibinfo {author} {\bibfnamefont {W.-G.}\ \bibnamefont {Yin}}, \bibinfo
  {author} {\bibfnamefont {W.}~\bibnamefont {Ku}}, \bibinfo {author}
  {\bibfnamefont {A.}~\bibnamefont {Tsvelik}}, \bibinfo {author} {\bibfnamefont
  {Y.-J.}\ \bibnamefont {Kim}}, \bibinfo {author} {\bibfnamefont
  {H.}~\bibnamefont {Gretarsson}}, \bibinfo {author} {\bibfnamefont
  {Y.}~\bibnamefont {Singh}}, \bibinfo {author} {\bibfnamefont
  {P.}~\bibnamefont {Gegenwart}}, \ and\ \bibinfo {author} {\bibfnamefont
  {J.~P.}\ \bibnamefont {Hill}},\ }\href {\doibase 10.1103/PhysRevB.83.220403}
  {\bibfield  {journal} {\bibinfo  {journal} {Phys. Rev. B}\ }\textbf {\bibinfo
  {volume} {83}},\ \bibinfo {pages} {220403} (\bibinfo {year}
  {2011})}\BibitemShut {NoStop}%
\bibitem [{\citenamefont {Choi}\ \emph {et~al.}(2012)\citenamefont {Choi},
  \citenamefont {Coldea}, \citenamefont {Kolmogorov}, \citenamefont
  {Lancaster}, \citenamefont {Mazin}, \citenamefont {Blundell}, \citenamefont
  {Radaelli}, \citenamefont {Singh}, \citenamefont {Gegenwart}, \citenamefont
  {Choi}, \citenamefont {Cheong}, \citenamefont {Baker}, \citenamefont
  {Stock},\ and\ \citenamefont {Taylor}}]{TaylorPRL2012}%
  \BibitemOpen
  \bibfield  {author} {\bibinfo {author} {\bibfnamefont {S.~K.}\ \bibnamefont
  {Choi}}, \bibinfo {author} {\bibfnamefont {R.}~\bibnamefont {Coldea}},
  \bibinfo {author} {\bibfnamefont {A.~N.}\ \bibnamefont {Kolmogorov}},
  \bibinfo {author} {\bibfnamefont {T.}~\bibnamefont {Lancaster}}, \bibinfo
  {author} {\bibfnamefont {I.~I.}\ \bibnamefont {Mazin}}, \bibinfo {author}
  {\bibfnamefont {S.~J.}\ \bibnamefont {Blundell}}, \bibinfo {author}
  {\bibfnamefont {P.~G.}\ \bibnamefont {Radaelli}}, \bibinfo {author}
  {\bibfnamefont {Y.}~\bibnamefont {Singh}}, \bibinfo {author} {\bibfnamefont
  {P.}~\bibnamefont {Gegenwart}}, \bibinfo {author} {\bibfnamefont {K.~R.}\
  \bibnamefont {Choi}}, \bibinfo {author} {\bibfnamefont {S.-W.}\ \bibnamefont
  {Cheong}}, \bibinfo {author} {\bibfnamefont {P.~J.}\ \bibnamefont {Baker}},
  \bibinfo {author} {\bibfnamefont {C.}~\bibnamefont {Stock}}, \ and\ \bibinfo
  {author} {\bibfnamefont {J.}~\bibnamefont {Taylor}},\ }\href {\doibase
  10.1103/PhysRevLett.108.127204} {\bibfield  {journal} {\bibinfo  {journal}
  {Phys. Rev. Lett.}\ }\textbf {\bibinfo {volume} {108}},\ \bibinfo {pages}
  {127204} (\bibinfo {year} {2012})}\BibitemShut {NoStop}%
\bibitem [{\citenamefont {Ye}\ \emph {et~al.}(2012)\citenamefont {Ye},
  \citenamefont {Chi}, \citenamefont {Cao}, \citenamefont {Chakoumakos},
  \citenamefont {Fernandez-Baca}, \citenamefont {Custelcean}, \citenamefont
  {Qi}, \citenamefont {Korneta},\ and\ \citenamefont {Cao}}]{CaoPRB2012}%
  \BibitemOpen
  \bibfield  {author} {\bibinfo {author} {\bibfnamefont {F.}~\bibnamefont
  {Ye}}, \bibinfo {author} {\bibfnamefont {S.}~\bibnamefont {Chi}}, \bibinfo
  {author} {\bibfnamefont {H.}~\bibnamefont {Cao}}, \bibinfo {author}
  {\bibfnamefont {B.~C.}\ \bibnamefont {Chakoumakos}}, \bibinfo {author}
  {\bibfnamefont {J.~A.}\ \bibnamefont {Fernandez-Baca}}, \bibinfo {author}
  {\bibfnamefont {R.}~\bibnamefont {Custelcean}}, \bibinfo {author}
  {\bibfnamefont {T.~F.}\ \bibnamefont {Qi}}, \bibinfo {author} {\bibfnamefont
  {O.~B.}\ \bibnamefont {Korneta}}, \ and\ \bibinfo {author} {\bibfnamefont
  {G.}~\bibnamefont {Cao}},\ }\href {\doibase 10.1103/PhysRevB.85.180403}
  {\bibfield  {journal} {\bibinfo  {journal} {Phys. Rev. B}\ }\textbf {\bibinfo
  {volume} {85}},\ \bibinfo {pages} {180403} (\bibinfo {year}
  {2012})}\BibitemShut {NoStop}%
\bibitem [{\citenamefont {Sears}\ \emph {et~al.}(2015)\citenamefont {Sears},
  \citenamefont {Songvilay}, \citenamefont {Plumb}, \citenamefont {Clancy},
  \citenamefont {Qiu}, \citenamefont {Zhao}, \citenamefont {Parshall},\ and\
  \citenamefont {Kim}}]{KimPRB2015}%
  \BibitemOpen
  \bibfield  {author} {\bibinfo {author} {\bibfnamefont {J.~A.}\ \bibnamefont
  {Sears}}, \bibinfo {author} {\bibfnamefont {M.}~\bibnamefont {Songvilay}},
  \bibinfo {author} {\bibfnamefont {K.~W.}\ \bibnamefont {Plumb}}, \bibinfo
  {author} {\bibfnamefont {J.~P.}\ \bibnamefont {Clancy}}, \bibinfo {author}
  {\bibfnamefont {Y.}~\bibnamefont {Qiu}}, \bibinfo {author} {\bibfnamefont
  {Y.}~\bibnamefont {Zhao}}, \bibinfo {author} {\bibfnamefont {D.}~\bibnamefont
  {Parshall}}, \ and\ \bibinfo {author} {\bibfnamefont {Y.-J.}\ \bibnamefont
  {Kim}},\ }\href {\doibase 10.1103/PhysRevB.91.144420} {\bibfield  {journal}
  {\bibinfo  {journal} {Phys. Rev. B}\ }\textbf {\bibinfo {volume} {91}},\
  \bibinfo {pages} {144420} (\bibinfo {year} {2015})}\BibitemShut {NoStop}%
\bibitem [{\citenamefont {Biffin}\ \emph
  {et~al.}(2014{\natexlab{a}})\citenamefont {Biffin}, \citenamefont {Johnson},
  \citenamefont {Choi}, \citenamefont {Freund}, \citenamefont {Manni},
  \citenamefont {Bombardi}, \citenamefont {Manuel}, \citenamefont {Gegenwart},\
  and\ \citenamefont {Coldea}}]{ColdeaPRB2014}%
  \BibitemOpen
  \bibfield  {author} {\bibinfo {author} {\bibfnamefont {A.}~\bibnamefont
  {Biffin}}, \bibinfo {author} {\bibfnamefont {R.~D.}\ \bibnamefont {Johnson}},
  \bibinfo {author} {\bibfnamefont {S.}~\bibnamefont {Choi}}, \bibinfo {author}
  {\bibfnamefont {F.}~\bibnamefont {Freund}}, \bibinfo {author} {\bibfnamefont
  {S.}~\bibnamefont {Manni}}, \bibinfo {author} {\bibfnamefont
  {A.}~\bibnamefont {Bombardi}}, \bibinfo {author} {\bibfnamefont
  {P.}~\bibnamefont {Manuel}}, \bibinfo {author} {\bibfnamefont
  {P.}~\bibnamefont {Gegenwart}}, \ and\ \bibinfo {author} {\bibfnamefont
  {R.}~\bibnamefont {Coldea}},\ }\href {\doibase 10.1103/PhysRevB.90.205116}
  {\bibfield  {journal} {\bibinfo  {journal} {Phys. Rev. B}\ }\textbf {\bibinfo
  {volume} {90}},\ \bibinfo {pages} {205116} (\bibinfo {year}
  {2014}{\natexlab{a}})}\BibitemShut {NoStop}%
\bibitem [{\citenamefont {Biffin}\ \emph
  {et~al.}(2014{\natexlab{b}})\citenamefont {Biffin}, \citenamefont {Johnson},
  \citenamefont {Kimchi}, \citenamefont {Morris}, \citenamefont {Bombardi},
  \citenamefont {Analytis}, \citenamefont {Vishwanath},\ and\ \citenamefont
  {Coldea}}]{ColdeaPRL2014}%
  \BibitemOpen
  \bibfield  {author} {\bibinfo {author} {\bibfnamefont {A.}~\bibnamefont
  {Biffin}}, \bibinfo {author} {\bibfnamefont {R.~D.}\ \bibnamefont {Johnson}},
  \bibinfo {author} {\bibfnamefont {I.}~\bibnamefont {Kimchi}}, \bibinfo
  {author} {\bibfnamefont {R.}~\bibnamefont {Morris}}, \bibinfo {author}
  {\bibfnamefont {A.}~\bibnamefont {Bombardi}}, \bibinfo {author}
  {\bibfnamefont {J.~G.}\ \bibnamefont {Analytis}}, \bibinfo {author}
  {\bibfnamefont {A.}~\bibnamefont {Vishwanath}}, \ and\ \bibinfo {author}
  {\bibfnamefont {R.}~\bibnamefont {Coldea}},\ }\href {\doibase
  10.1103/PhysRevLett.113.197201} {\bibfield  {journal} {\bibinfo  {journal}
  {Phys. Rev. Lett.}\ }\textbf {\bibinfo {volume} {113}},\ \bibinfo {pages}
  {197201} (\bibinfo {year} {2014}{\natexlab{b}})}\BibitemShut {NoStop}%
\bibitem [{\citenamefont {Williams}\ \emph {et~al.}(2016)\citenamefont
  {Williams}, \citenamefont {Johnson}, \citenamefont {Freund}, \citenamefont
  {Choi}, \citenamefont {Jesche}, \citenamefont {Kimchi}, \citenamefont
  {Manni}, \citenamefont {Bombardi}, \citenamefont {Manuel}, \citenamefont
  {Gegenwart},\ and\ \citenamefont {Coldea}}]{ColdeaPRB2016}%
  \BibitemOpen
  \bibfield  {author} {\bibinfo {author} {\bibfnamefont {S.~C.}\ \bibnamefont
  {Williams}}, \bibinfo {author} {\bibfnamefont {R.~D.}\ \bibnamefont
  {Johnson}}, \bibinfo {author} {\bibfnamefont {F.}~\bibnamefont {Freund}},
  \bibinfo {author} {\bibfnamefont {S.}~\bibnamefont {Choi}}, \bibinfo {author}
  {\bibfnamefont {A.}~\bibnamefont {Jesche}}, \bibinfo {author} {\bibfnamefont
  {I.}~\bibnamefont {Kimchi}}, \bibinfo {author} {\bibfnamefont
  {S.}~\bibnamefont {Manni}}, \bibinfo {author} {\bibfnamefont
  {A.}~\bibnamefont {Bombardi}}, \bibinfo {author} {\bibfnamefont
  {P.}~\bibnamefont {Manuel}}, \bibinfo {author} {\bibfnamefont
  {P.}~\bibnamefont {Gegenwart}}, \ and\ \bibinfo {author} {\bibfnamefont
  {R.}~\bibnamefont {Coldea}},\ }\href {\doibase 10.1103/PhysRevB.93.195158}
  {\bibfield  {journal} {\bibinfo  {journal} {Phys. Rev. B}\ }\textbf {\bibinfo
  {volume} {93}},\ \bibinfo {pages} {195158} (\bibinfo {year}
  {2016})}\BibitemShut {NoStop}%
\bibitem [{\citenamefont {Chaloupka}\ \emph {et~al.}(2010)\citenamefont
  {Chaloupka}, \citenamefont {Jackeli},\ and\ \citenamefont
  {Khaliullin}}]{KhaliullinPRL2010}%
  \BibitemOpen
  \bibfield  {author} {\bibinfo {author} {\bibfnamefont {J.}~\bibnamefont
  {Chaloupka}}, \bibinfo {author} {\bibfnamefont {G.}~\bibnamefont {Jackeli}},
  \ and\ \bibinfo {author} {\bibfnamefont {G.}~\bibnamefont {Khaliullin}},\
  }\href {\doibase 10.1103/PhysRevLett.105.027204} {\bibfield  {journal}
  {\bibinfo  {journal} {Phys. Rev. Lett.}\ }\textbf {\bibinfo {volume} {105}},\
  \bibinfo {pages} {027204} (\bibinfo {year} {2010})}\BibitemShut {NoStop}%
\bibitem [{\citenamefont {Reuther}\ \emph {et~al.}(2011)\citenamefont
  {Reuther}, \citenamefont {Thomale},\ and\ \citenamefont
  {Trebst}}]{TrebstPRB2011}%
  \BibitemOpen
  \bibfield  {author} {\bibinfo {author} {\bibfnamefont {J.}~\bibnamefont
  {Reuther}}, \bibinfo {author} {\bibfnamefont {R.}~\bibnamefont {Thomale}}, \
  and\ \bibinfo {author} {\bibfnamefont {S.}~\bibnamefont {Trebst}},\ }\href
  {\doibase 10.1103/PhysRevB.84.100406} {\bibfield  {journal} {\bibinfo
  {journal} {Phys. Rev. B}\ }\textbf {\bibinfo {volume} {84}},\ \bibinfo
  {pages} {100406} (\bibinfo {year} {2011})}\BibitemShut {NoStop}%
\bibitem [{\citenamefont {Chaloupka}\ \emph {et~al.}(2013)\citenamefont
  {Chaloupka}, \citenamefont {Jackeli},\ and\ \citenamefont
  {Khaliullin}}]{ChaloupkaPRL2013}%
  \BibitemOpen
  \bibfield  {author} {\bibinfo {author} {\bibfnamefont {J.}~\bibnamefont
  {Chaloupka}}, \bibinfo {author} {\bibfnamefont {G.}~\bibnamefont {Jackeli}},
  \ and\ \bibinfo {author} {\bibfnamefont {G.}~\bibnamefont {Khaliullin}},\
  }\href {\doibase 10.1103/PhysRevLett.110.097204} {\bibfield  {journal}
  {\bibinfo  {journal} {Phys. Rev. Lett.}\ }\textbf {\bibinfo {volume} {110}},\
  \bibinfo {pages} {097204} (\bibinfo {year} {2013})}\BibitemShut {NoStop}%
\bibitem [{\citenamefont {Sizyuk}\ \emph {et~al.}(2014)\citenamefont {Sizyuk},
  \citenamefont {Price}, \citenamefont {W\"olfle},\ and\ \citenamefont
  {Perkins}}]{PerkinsPRB2014}%
  \BibitemOpen
  \bibfield  {author} {\bibinfo {author} {\bibfnamefont {Y.}~\bibnamefont
  {Sizyuk}}, \bibinfo {author} {\bibfnamefont {C.}~\bibnamefont {Price}},
  \bibinfo {author} {\bibfnamefont {P.}~\bibnamefont {W\"olfle}}, \ and\
  \bibinfo {author} {\bibfnamefont {N.~B.}\ \bibnamefont {Perkins}},\ }\href
  {\doibase 10.1103/PhysRevB.90.155126} {\bibfield  {journal} {\bibinfo
  {journal} {Phys. Rev. B}\ }\textbf {\bibinfo {volume} {90}},\ \bibinfo
  {pages} {155126} (\bibinfo {year} {2014})}\BibitemShut {NoStop}%
\bibitem [{\citenamefont {Rau}\ \emph {et~al.}(2014)\citenamefont {Rau},
  \citenamefont {Lee},\ and\ \citenamefont {Kee}}]{RauPRL2014}%
  \BibitemOpen
  \bibfield  {author} {\bibinfo {author} {\bibfnamefont {J.~G.}\ \bibnamefont
  {Rau}}, \bibinfo {author} {\bibfnamefont {E.~K.-H.}\ \bibnamefont {Lee}}, \
  and\ \bibinfo {author} {\bibfnamefont {H.-Y.}\ \bibnamefont {Kee}},\ }\href
  {\doibase 10.1103/PhysRevLett.112.077204} {\bibfield  {journal} {\bibinfo
  {journal} {Phys. Rev. Lett.}\ }\textbf {\bibinfo {volume} {112}},\ \bibinfo
  {pages} {077204} (\bibinfo {year} {2014})}\BibitemShut {NoStop}%
\bibitem [{\citenamefont {Yamaji}\ \emph {et~al.}(2014)\citenamefont {Yamaji},
  \citenamefont {Nomura}, \citenamefont {Kurita}, \citenamefont {Arita},\ and\
  \citenamefont {Imada}}]{ImadaPRL2014}%
  \BibitemOpen
  \bibfield  {author} {\bibinfo {author} {\bibfnamefont {Y.}~\bibnamefont
  {Yamaji}}, \bibinfo {author} {\bibfnamefont {Y.}~\bibnamefont {Nomura}},
  \bibinfo {author} {\bibfnamefont {M.}~\bibnamefont {Kurita}}, \bibinfo
  {author} {\bibfnamefont {R.}~\bibnamefont {Arita}}, \ and\ \bibinfo {author}
  {\bibfnamefont {M.}~\bibnamefont {Imada}},\ }\href {\doibase
  10.1103/PhysRevLett.113.107201} {\bibfield  {journal} {\bibinfo  {journal}
  {Phys. Rev. Lett.}\ }\textbf {\bibinfo {volume} {113}},\ \bibinfo {pages}
  {107201} (\bibinfo {year} {2014})}\BibitemShut {NoStop}%
\bibitem [{\citenamefont {{Katukuri}}\ \emph {et~al.}(2014)\citenamefont
  {{Katukuri}}, \citenamefont {{Nishimoto}}, \citenamefont {{Yushankhai}},
  \citenamefont {{Stoyanova}}, \citenamefont {{Kandpal}}, \citenamefont
  {{Choi}}, \citenamefont {{Coldea}}, \citenamefont {{Rousochatzakis}},
  \citenamefont {{Hozoi}},\ and\ \citenamefont {{van den
  Brink}}}]{VanderBrinkNJoP2014}%
  \BibitemOpen
  \bibfield  {author} {\bibinfo {author} {\bibfnamefont {V.~M.}\ \bibnamefont
  {{Katukuri}}}, \bibinfo {author} {\bibfnamefont {S.}~\bibnamefont
  {{Nishimoto}}}, \bibinfo {author} {\bibfnamefont {V.}~\bibnamefont
  {{Yushankhai}}}, \bibinfo {author} {\bibfnamefont {A.}~\bibnamefont
  {{Stoyanova}}}, \bibinfo {author} {\bibfnamefont {H.}~\bibnamefont
  {{Kandpal}}}, \bibinfo {author} {\bibfnamefont {S.}~\bibnamefont {{Choi}}},
  \bibinfo {author} {\bibfnamefont {R.}~\bibnamefont {{Coldea}}}, \bibinfo
  {author} {\bibfnamefont {I.}~\bibnamefont {{Rousochatzakis}}}, \bibinfo
  {author} {\bibfnamefont {L.}~\bibnamefont {{Hozoi}}}, \ and\ \bibinfo
  {author} {\bibfnamefont {J.}~\bibnamefont {{van den Brink}}},\ }\href
  {\doibase 10.1088/1367-2630/16/1/013056} {\bibfield  {journal} {\bibinfo
  {journal} {New Journal of Physics}\ }\textbf {\bibinfo {volume} {16}},\
  \bibinfo {eid} {013056} (\bibinfo {year} {2014})}\BibitemShut {NoStop}%
\bibitem [{\citenamefont {Witczak-Krempa}\ \emph {et~al.}(2014)\citenamefont
  {Witczak-Krempa}, \citenamefont {Chen}, \citenamefont {Kim},\ and\
  \citenamefont {Balents}}]{KrempaARCMP2014}%
  \BibitemOpen
  \bibfield  {author} {\bibinfo {author} {\bibfnamefont {W.}~\bibnamefont
  {Witczak-Krempa}}, \bibinfo {author} {\bibfnamefont {G.}~\bibnamefont
  {Chen}}, \bibinfo {author} {\bibfnamefont {Y.~B.}\ \bibnamefont {Kim}}, \
  and\ \bibinfo {author} {\bibfnamefont {L.}~\bibnamefont {Balents}},\ }\href
  {https://doi.org/10.1146/annurev-conmatphys-020911-125138} {\bibfield
  {journal} {\bibinfo  {journal} {Annual Review of Condensed Matter Physics}\
  }\textbf {\bibinfo {volume} {5}},\ \bibinfo {pages} {57} (\bibinfo {year}
  {2014})}\BibitemShut {NoStop}%
\bibitem [{\citenamefont {Rau}\ \emph {et~al.}(2016)\citenamefont {Rau},
  \citenamefont {Lee},\ and\ \citenamefont {Kee}}]{RauARCMP2016}%
  \BibitemOpen
  \bibfield  {author} {\bibinfo {author} {\bibfnamefont {J.~G.}\ \bibnamefont
  {Rau}}, \bibinfo {author} {\bibfnamefont {E.~K.-H.}\ \bibnamefont {Lee}}, \
  and\ \bibinfo {author} {\bibfnamefont {H.-Y.}\ \bibnamefont {Kee}},\ }\href
  {https://doi.org/10.1146/annurev-conmatphys-031115-011319} {\bibfield
  {journal} {\bibinfo  {journal} {Annual Review of Condensed Matter Physics}\
  }\textbf {\bibinfo {volume} {7}},\ \bibinfo {pages} {195} (\bibinfo {year}
  {2016})}\BibitemShut {NoStop}%
\bibitem [{\citenamefont {Katukuri}\ \emph {et~al.}(2016)\citenamefont
  {Katukuri}, \citenamefont {Yadav}, \citenamefont {Hozoi}, \citenamefont
  {Nishimoto},\ and\ \citenamefont {van~den Brink}}]{VanDenBrinkSR2016}%
  \BibitemOpen
  \bibfield  {author} {\bibinfo {author} {\bibfnamefont {V.~M.}\ \bibnamefont
  {Katukuri}}, \bibinfo {author} {\bibfnamefont {R.}~\bibnamefont {Yadav}},
  \bibinfo {author} {\bibfnamefont {L.}~\bibnamefont {Hozoi}}, \bibinfo
  {author} {\bibfnamefont {S.}~\bibnamefont {Nishimoto}}, \ and\ \bibinfo
  {author} {\bibfnamefont {J.}~\bibnamefont {van~den Brink}},\ }\href
  {https://www.nature.com/articles/srep29585} {\bibfield  {journal} {\bibinfo
  {journal} {Scientific Reports}\ }\textbf {\bibinfo {volume} {6}},\ \bibinfo
  {pages} {29585} (\bibinfo {year} {2016})}\BibitemShut {NoStop}%
\bibitem [{\citenamefont {Kim}\ \emph {et~al.}(2015{\natexlab{b}})\citenamefont
  {Kim}, \citenamefont {Lee},\ and\ \citenamefont {Kim}}]{KimEPL2015}%
  \BibitemOpen
  \bibfield  {author} {\bibinfo {author} {\bibfnamefont {H.-S.}\ \bibnamefont
  {Kim}}, \bibinfo {author} {\bibfnamefont {E.~K.-H.}\ \bibnamefont {Lee}}, \
  and\ \bibinfo {author} {\bibfnamefont {Y.~B.}\ \bibnamefont {Kim}},\ }\href
  {http://stacks.iop.org/0295-5075/112/i=6/a=67004} {\bibfield  {journal}
  {\bibinfo  {journal} {EPL (Europhysics Letters)}\ }\textbf {\bibinfo {volume}
  {112}},\ \bibinfo {pages} {67004} (\bibinfo {year}
  {2015}{\natexlab{b}})}\BibitemShut {NoStop}%
\bibitem [{\citenamefont {Kim}\ \emph {et~al.}(2016)\citenamefont {Kim},
  \citenamefont {Kim},\ and\ \citenamefont {Kee}}]{KimPRB2016}%
  \BibitemOpen
  \bibfield  {author} {\bibinfo {author} {\bibfnamefont {H.-S.}\ \bibnamefont
  {Kim}}, \bibinfo {author} {\bibfnamefont {Y.~B.}\ \bibnamefont {Kim}}, \ and\
  \bibinfo {author} {\bibfnamefont {H.-Y.}\ \bibnamefont {Kee}},\ }\href
  {\doibase 10.1103/PhysRevB.94.245127} {\bibfield  {journal} {\bibinfo
  {journal} {Phys. Rev. B}\ }\textbf {\bibinfo {volume} {94}},\ \bibinfo
  {pages} {245127} (\bibinfo {year} {2016})}\BibitemShut {NoStop}%
\bibitem [{\citenamefont {Lee}\ and\ \citenamefont {Kim}(2015)}]{LeePRB2015}%
  \BibitemOpen
  \bibfield  {author} {\bibinfo {author} {\bibfnamefont {E.~K.-H.}\
  \bibnamefont {Lee}}\ and\ \bibinfo {author} {\bibfnamefont {Y.~B.}\
  \bibnamefont {Kim}},\ }\href {\doibase 10.1103/PhysRevB.91.064407} {\bibfield
   {journal} {\bibinfo  {journal} {Phys. Rev. B}\ }\textbf {\bibinfo {volume}
  {91}},\ \bibinfo {pages} {064407} (\bibinfo {year} {2015})}\BibitemShut
  {NoStop}%
\bibitem [{\citenamefont {Chaloupka}\ and\ \citenamefont
  {Khaliullin}(2015)}]{KhaliullinPRB2015}%
  \BibitemOpen
  \bibfield  {author} {\bibinfo {author} {\bibfnamefont {J.}~\bibnamefont
  {Chaloupka}}\ and\ \bibinfo {author} {\bibfnamefont {G.}~\bibnamefont
  {Khaliullin}},\ }\href {\doibase 10.1103/PhysRevB.92.024413} {\bibfield
  {journal} {\bibinfo  {journal} {Phys. Rev. B}\ }\textbf {\bibinfo {volume}
  {92}},\ \bibinfo {pages} {024413} (\bibinfo {year} {2015})}\BibitemShut
  {NoStop}%
\bibitem [{\citenamefont {Ducatman}\ \emph {et~al.}(2018)\citenamefont
  {Ducatman}, \citenamefont {Rousochatzakis},\ and\ \citenamefont
  {Perkins}}]{NataliaPRB2018}%
  \BibitemOpen
  \bibfield  {author} {\bibinfo {author} {\bibfnamefont {S.}~\bibnamefont
  {Ducatman}}, \bibinfo {author} {\bibfnamefont {I.}~\bibnamefont
  {Rousochatzakis}}, \ and\ \bibinfo {author} {\bibfnamefont {N.~B.}\
  \bibnamefont {Perkins}},\ }\href {\doibase 10.1103/PhysRevB.97.125125}
  {\bibfield  {journal} {\bibinfo  {journal} {Phys. Rev. B}\ }\textbf {\bibinfo
  {volume} {97}},\ \bibinfo {pages} {125125} (\bibinfo {year}
  {2018})}\BibitemShut {NoStop}%
\bibitem [{\citenamefont {Kim}\ \emph {et~al.}(2008)\citenamefont {Kim},
  \citenamefont {Jin}, \citenamefont {Moon}, \citenamefont {Kim}, \citenamefont
  {Park}, \citenamefont {Leem}, \citenamefont {Yu}, \citenamefont {Noh},
  \citenamefont {Kim}, \citenamefont {Oh}, \citenamefont {Park}, \citenamefont
  {Durairaj}, \citenamefont {Cao},\ and\ \citenamefont
  {Rotenberg}}]{RotenbergPRL2008}%
  \BibitemOpen
  \bibfield  {author} {\bibinfo {author} {\bibfnamefont {B.~J.}\ \bibnamefont
  {Kim}}, \bibinfo {author} {\bibfnamefont {H.}~\bibnamefont {Jin}}, \bibinfo
  {author} {\bibfnamefont {S.~J.}\ \bibnamefont {Moon}}, \bibinfo {author}
  {\bibfnamefont {J.-Y.}\ \bibnamefont {Kim}}, \bibinfo {author} {\bibfnamefont
  {B.-G.}\ \bibnamefont {Park}}, \bibinfo {author} {\bibfnamefont {C.~S.}\
  \bibnamefont {Leem}}, \bibinfo {author} {\bibfnamefont {J.}~\bibnamefont
  {Yu}}, \bibinfo {author} {\bibfnamefont {T.~W.}\ \bibnamefont {Noh}},
  \bibinfo {author} {\bibfnamefont {C.}~\bibnamefont {Kim}}, \bibinfo {author}
  {\bibfnamefont {S.-J.}\ \bibnamefont {Oh}}, \bibinfo {author} {\bibfnamefont
  {J.-H.}\ \bibnamefont {Park}}, \bibinfo {author} {\bibfnamefont
  {V.}~\bibnamefont {Durairaj}}, \bibinfo {author} {\bibfnamefont
  {G.}~\bibnamefont {Cao}}, \ and\ \bibinfo {author} {\bibfnamefont
  {E.}~\bibnamefont {Rotenberg}},\ }\href {\doibase
  10.1103/PhysRevLett.101.076402} {\bibfield  {journal} {\bibinfo  {journal}
  {Phys. Rev. Lett.}\ }\textbf {\bibinfo {volume} {101}},\ \bibinfo {pages}
  {076402} (\bibinfo {year} {2008})}\BibitemShut {NoStop}%
\bibitem [{\citenamefont {Gretarsson}\ \emph {et~al.}(2013)\citenamefont
  {Gretarsson}, \citenamefont {Clancy}, \citenamefont {Liu}, \citenamefont
  {Hill}, \citenamefont {Bozin}, \citenamefont {Singh}, \citenamefont {Manni},
  \citenamefont {Gegenwart}, \citenamefont {Kim}, \citenamefont {Said},
  \citenamefont {Casa}, \citenamefont {Gog}, \citenamefont {Upton},
  \citenamefont {Kim}, \citenamefont {Yu}, \citenamefont {Katukuri},
  \citenamefont {Hozoi}, \citenamefont {van~den Brink},\ and\ \citenamefont
  {Kim}}]{GretarssonPRL2013}%
  \BibitemOpen
  \bibfield  {author} {\bibinfo {author} {\bibfnamefont {H.}~\bibnamefont
  {Gretarsson}}, \bibinfo {author} {\bibfnamefont {J.~P.}\ \bibnamefont
  {Clancy}}, \bibinfo {author} {\bibfnamefont {X.}~\bibnamefont {Liu}},
  \bibinfo {author} {\bibfnamefont {J.~P.}\ \bibnamefont {Hill}}, \bibinfo
  {author} {\bibfnamefont {E.}~\bibnamefont {Bozin}}, \bibinfo {author}
  {\bibfnamefont {Y.}~\bibnamefont {Singh}}, \bibinfo {author} {\bibfnamefont
  {S.}~\bibnamefont {Manni}}, \bibinfo {author} {\bibfnamefont
  {P.}~\bibnamefont {Gegenwart}}, \bibinfo {author} {\bibfnamefont
  {J.}~\bibnamefont {Kim}}, \bibinfo {author} {\bibfnamefont {A.~H.}\
  \bibnamefont {Said}}, \bibinfo {author} {\bibfnamefont {D.}~\bibnamefont
  {Casa}}, \bibinfo {author} {\bibfnamefont {T.}~\bibnamefont {Gog}}, \bibinfo
  {author} {\bibfnamefont {M.~H.}\ \bibnamefont {Upton}}, \bibinfo {author}
  {\bibfnamefont {H.-S.}\ \bibnamefont {Kim}}, \bibinfo {author} {\bibfnamefont
  {J.}~\bibnamefont {Yu}}, \bibinfo {author} {\bibfnamefont {V.~M.}\
  \bibnamefont {Katukuri}}, \bibinfo {author} {\bibfnamefont {L.}~\bibnamefont
  {Hozoi}}, \bibinfo {author} {\bibfnamefont {J.}~\bibnamefont {van~den
  Brink}}, \ and\ \bibinfo {author} {\bibfnamefont {Y.-J.}\ \bibnamefont
  {Kim}},\ }\href {\doibase 10.1103/PhysRevLett.110.076402} {\bibfield
  {journal} {\bibinfo  {journal} {Phys. Rev. Lett.}\ }\textbf {\bibinfo
  {volume} {110}},\ \bibinfo {pages} {076402} (\bibinfo {year}
  {2013})}\BibitemShut {NoStop}%
\bibitem [{\citenamefont {Mandal}\ and\ \citenamefont
  {Surendran}(2009)}]{SurendranPRB2009}%
  \BibitemOpen
  \bibfield  {author} {\bibinfo {author} {\bibfnamefont {S.}~\bibnamefont
  {Mandal}}\ and\ \bibinfo {author} {\bibfnamefont {N.}~\bibnamefont
  {Surendran}},\ }\href {\doibase 10.1103/PhysRevB.79.024426} {\bibfield
  {journal} {\bibinfo  {journal} {Phys. Rev. B}\ }\textbf {\bibinfo {volume}
  {79}},\ \bibinfo {pages} {024426} (\bibinfo {year} {2009})}\BibitemShut
  {NoStop}%
\bibitem [{\citenamefont {Kimchi}\ \emph {et~al.}(2014)\citenamefont {Kimchi},
  \citenamefont {Analytis},\ and\ \citenamefont {Vishwanath}}]{KimchiPRB2014}%
  \BibitemOpen
  \bibfield  {author} {\bibinfo {author} {\bibfnamefont {I.}~\bibnamefont
  {Kimchi}}, \bibinfo {author} {\bibfnamefont {J.~G.}\ \bibnamefont
  {Analytis}}, \ and\ \bibinfo {author} {\bibfnamefont {A.}~\bibnamefont
  {Vishwanath}},\ }\href {\doibase 10.1103/PhysRevB.90.205126} {\bibfield
  {journal} {\bibinfo  {journal} {Phys. Rev. B}\ }\textbf {\bibinfo {volume}
  {90}},\ \bibinfo {pages} {205126} (\bibinfo {year} {2014})}\BibitemShut
  {NoStop}%
\bibitem [{\citenamefont {Lee}\ \emph {et~al.}(2014)\citenamefont {Lee},
  \citenamefont {Schaffer}, \citenamefont {Bhattacharjee},\ and\ \citenamefont
  {Kim}}]{LeePRB2014}%
  \BibitemOpen
  \bibfield  {author} {\bibinfo {author} {\bibfnamefont {E.~K.-H.}\
  \bibnamefont {Lee}}, \bibinfo {author} {\bibfnamefont {R.}~\bibnamefont
  {Schaffer}}, \bibinfo {author} {\bibfnamefont {S.}~\bibnamefont
  {Bhattacharjee}}, \ and\ \bibinfo {author} {\bibfnamefont {Y.~B.}\
  \bibnamefont {Kim}},\ }\href {\doibase 10.1103/PhysRevB.89.045117} {\bibfield
   {journal} {\bibinfo  {journal} {Phys. Rev. B}\ }\textbf {\bibinfo {volume}
  {89}},\ \bibinfo {pages} {045117} (\bibinfo {year} {2014})}\BibitemShut
  {NoStop}%
\bibitem [{\citenamefont {Schaffer}\ \emph {et~al.}(2015)\citenamefont
  {Schaffer}, \citenamefont {Lee}, \citenamefont {Lu},\ and\ \citenamefont
  {Kim}}]{SchafferPRL2015}%
  \BibitemOpen
  \bibfield  {author} {\bibinfo {author} {\bibfnamefont {R.}~\bibnamefont
  {Schaffer}}, \bibinfo {author} {\bibfnamefont {E.~K.-H.}\ \bibnamefont
  {Lee}}, \bibinfo {author} {\bibfnamefont {Y.-M.}\ \bibnamefont {Lu}}, \ and\
  \bibinfo {author} {\bibfnamefont {Y.~B.}\ \bibnamefont {Kim}},\ }\href
  {\doibase 10.1103/PhysRevLett.114.116803} {\bibfield  {journal} {\bibinfo
  {journal} {Phys. Rev. Lett.}\ }\textbf {\bibinfo {volume} {114}},\ \bibinfo
  {pages} {116803} (\bibinfo {year} {2015})}\BibitemShut {NoStop}%
\bibitem [{\citenamefont {Chaloupka}\ and\ \citenamefont
  {Khaliullin}(2016)}]{KhaliullinPRB2016}%
  \BibitemOpen
  \bibfield  {author} {\bibinfo {author} {\bibfnamefont {J.}~\bibnamefont
  {Chaloupka}}\ and\ \bibinfo {author} {\bibfnamefont {G.}~\bibnamefont
  {Khaliullin}},\ }\href {\doibase 10.1103/PhysRevB.94.064435} {\bibfield
  {journal} {\bibinfo  {journal} {Phys. Rev. B}\ }\textbf {\bibinfo {volume}
  {94}},\ \bibinfo {pages} {064435} (\bibinfo {year} {2016})}\BibitemShut
  {NoStop}%
\bibitem [{\citenamefont {Feng}\ \emph {et~al.}(2007)\citenamefont {Feng},
  \citenamefont {Zhang},\ and\ \citenamefont {Xiang}}]{XiangPRL2007}%
  \BibitemOpen
  \bibfield  {author} {\bibinfo {author} {\bibfnamefont {X.-Y.}\ \bibnamefont
  {Feng}}, \bibinfo {author} {\bibfnamefont {G.-M.}\ \bibnamefont {Zhang}}, \
  and\ \bibinfo {author} {\bibfnamefont {T.}~\bibnamefont {Xiang}},\ }\href
  {\doibase 10.1103/PhysRevLett.98.087204} {\bibfield  {journal} {\bibinfo
  {journal} {Phys. Rev. Lett.}\ }\textbf {\bibinfo {volume} {98}},\ \bibinfo
  {pages} {087204} (\bibinfo {year} {2007})}\BibitemShut {NoStop}%
\bibitem [{\citenamefont {Catuneanu}\ \emph {et~al.}(2018)\citenamefont
  {Catuneanu}, \citenamefont {Yamaji}, \citenamefont {Wachtel}, \citenamefont
  {Kim},\ and\ \citenamefont {Kee}}]{AndreiNPJ2018}%
  \BibitemOpen
  \bibfield  {author} {\bibinfo {author} {\bibfnamefont {A.}~\bibnamefont
  {Catuneanu}}, \bibinfo {author} {\bibfnamefont {Y.}~\bibnamefont {Yamaji}},
  \bibinfo {author} {\bibfnamefont {G.}~\bibnamefont {Wachtel}}, \bibinfo
  {author} {\bibfnamefont {Y.~B.}\ \bibnamefont {Kim}}, \ and\ \bibinfo
  {author} {\bibfnamefont {H.-Y.}\ \bibnamefont {Kee}},\ }\href {\doibase
  10.1038/s41535-018-0095-2} {\bibfield  {journal} {\bibinfo  {journal} {npj
  Quantum Materials}\ }\textbf {\bibinfo {volume} {3}},\ \bibinfo {pages} {23}
  (\bibinfo {year} {2018})}\BibitemShut {NoStop}%
\bibitem [{\citenamefont {Kimchi}\ \emph {et~al.}(2015)\citenamefont {Kimchi},
  \citenamefont {Coldea},\ and\ \citenamefont
  {Vishwanath}}]{VishwanathPRB2015}%
  \BibitemOpen
  \bibfield  {author} {\bibinfo {author} {\bibfnamefont {I.}~\bibnamefont
  {Kimchi}}, \bibinfo {author} {\bibfnamefont {R.}~\bibnamefont {Coldea}}, \
  and\ \bibinfo {author} {\bibfnamefont {A.}~\bibnamefont {Vishwanath}},\
  }\href {\doibase 10.1103/PhysRevB.91.245134} {\bibfield  {journal} {\bibinfo
  {journal} {Phys. Rev. B}\ }\textbf {\bibinfo {volume} {91}},\ \bibinfo
  {pages} {245134} (\bibinfo {year} {2015})}\BibitemShut {NoStop}%
\bibitem [{\citenamefont {Veiga}\ \emph {et~al.}(2017)\citenamefont {Veiga},
  \citenamefont {Etter}, \citenamefont {Glazyrin}, \citenamefont {Sun},
  \citenamefont {Escanhoela}, \citenamefont {Fabbris}, \citenamefont
  {Mardegan}, \citenamefont {Malavi}, \citenamefont {Deng}, \citenamefont
  {Stavropoulos}, \citenamefont {Kee}, \citenamefont {Yang}, \citenamefont {van
  Veenendaal}, \citenamefont {Schilling}, \citenamefont {Takayama},
  \citenamefont {Takagi},\ and\ \citenamefont {Haskel}}]{VeigaPRB2017}%
  \BibitemOpen
  \bibfield  {author} {\bibinfo {author} {\bibfnamefont {L.~S.~I.}\
  \bibnamefont {Veiga}}, \bibinfo {author} {\bibfnamefont {M.}~\bibnamefont
  {Etter}}, \bibinfo {author} {\bibfnamefont {K.}~\bibnamefont {Glazyrin}},
  \bibinfo {author} {\bibfnamefont {F.}~\bibnamefont {Sun}}, \bibinfo {author}
  {\bibfnamefont {C.~A.}\ \bibnamefont {Escanhoela}}, \bibinfo {author}
  {\bibfnamefont {G.}~\bibnamefont {Fabbris}}, \bibinfo {author} {\bibfnamefont
  {J.~R.~L.}\ \bibnamefont {Mardegan}}, \bibinfo {author} {\bibfnamefont
  {P.~S.}\ \bibnamefont {Malavi}}, \bibinfo {author} {\bibfnamefont
  {Y.}~\bibnamefont {Deng}}, \bibinfo {author} {\bibfnamefont {P.~P.}\
  \bibnamefont {Stavropoulos}}, \bibinfo {author} {\bibfnamefont {H.-Y.}\
  \bibnamefont {Kee}}, \bibinfo {author} {\bibfnamefont {W.~G.}\ \bibnamefont
  {Yang}}, \bibinfo {author} {\bibfnamefont {M.}~\bibnamefont {van
  Veenendaal}}, \bibinfo {author} {\bibfnamefont {J.~S.}\ \bibnamefont
  {Schilling}}, \bibinfo {author} {\bibfnamefont {T.}~\bibnamefont {Takayama}},
  \bibinfo {author} {\bibfnamefont {H.}~\bibnamefont {Takagi}}, \ and\ \bibinfo
  {author} {\bibfnamefont {D.}~\bibnamefont {Haskel}},\ }\href {\doibase
  10.1103/PhysRevB.96.140402} {\bibfield  {journal} {\bibinfo  {journal} {Phys.
  Rev. B}\ }\textbf {\bibinfo {volume} {96}},\ \bibinfo {pages} {140402}
  (\bibinfo {year} {2017})}\BibitemShut {NoStop}%
\bibitem [{\citenamefont {Breznay}\ \emph {et~al.}(2017)\citenamefont
  {Breznay}, \citenamefont {Ruiz}, \citenamefont {Frano}, \citenamefont {Bi},
  \citenamefont {Birgeneau}, \citenamefont {Haskel},\ and\ \citenamefont
  {Analytis}}]{BreznayPRB2017}%
  \BibitemOpen
  \bibfield  {author} {\bibinfo {author} {\bibfnamefont {N.~P.}\ \bibnamefont
  {Breznay}}, \bibinfo {author} {\bibfnamefont {A.}~\bibnamefont {Ruiz}},
  \bibinfo {author} {\bibfnamefont {A.}~\bibnamefont {Frano}}, \bibinfo
  {author} {\bibfnamefont {W.}~\bibnamefont {Bi}}, \bibinfo {author}
  {\bibfnamefont {R.~J.}\ \bibnamefont {Birgeneau}}, \bibinfo {author}
  {\bibfnamefont {D.}~\bibnamefont {Haskel}}, \ and\ \bibinfo {author}
  {\bibfnamefont {J.~G.}\ \bibnamefont {Analytis}},\ }\href {\doibase
  10.1103/PhysRevB.96.020402} {\bibfield  {journal} {\bibinfo  {journal} {Phys.
  Rev. B}\ }\textbf {\bibinfo {volume} {96}},\ \bibinfo {pages} {020402}
  (\bibinfo {year} {2017})}\BibitemShut {NoStop}%
\bibitem [{\citenamefont {Majumder}\ \emph {et~al.}(2018)\citenamefont
  {Majumder}, \citenamefont {Manna}, \citenamefont {Simutis}, \citenamefont
  {Orain}, \citenamefont {Dey}, \citenamefont {Freund}, \citenamefont {Jesche},
  \citenamefont {Khasanov}, \citenamefont {Biswas}, \citenamefont {Bykova},
  \citenamefont {Dubrovinskaia}, \citenamefont {Dubrovinsky}, \citenamefont
  {Yadav}, \citenamefont {Hozoi}, \citenamefont {Nishimoto}, \citenamefont
  {Tsirlin},\ and\ \citenamefont {Gegenwart}}]{GegenwartPRL2018}%
  \BibitemOpen
  \bibfield  {author} {\bibinfo {author} {\bibfnamefont {M.}~\bibnamefont
  {Majumder}}, \bibinfo {author} {\bibfnamefont {R.~S.}\ \bibnamefont {Manna}},
  \bibinfo {author} {\bibfnamefont {G.}~\bibnamefont {Simutis}}, \bibinfo
  {author} {\bibfnamefont {J.~C.}\ \bibnamefont {Orain}}, \bibinfo {author}
  {\bibfnamefont {T.}~\bibnamefont {Dey}}, \bibinfo {author} {\bibfnamefont
  {F.}~\bibnamefont {Freund}}, \bibinfo {author} {\bibfnamefont
  {A.}~\bibnamefont {Jesche}}, \bibinfo {author} {\bibfnamefont
  {R.}~\bibnamefont {Khasanov}}, \bibinfo {author} {\bibfnamefont {P.~K.}\
  \bibnamefont {Biswas}}, \bibinfo {author} {\bibfnamefont {E.}~\bibnamefont
  {Bykova}}, \bibinfo {author} {\bibfnamefont {N.}~\bibnamefont
  {Dubrovinskaia}}, \bibinfo {author} {\bibfnamefont {L.~S.}\ \bibnamefont
  {Dubrovinsky}}, \bibinfo {author} {\bibfnamefont {R.}~\bibnamefont {Yadav}},
  \bibinfo {author} {\bibfnamefont {L.}~\bibnamefont {Hozoi}}, \bibinfo
  {author} {\bibfnamefont {S.}~\bibnamefont {Nishimoto}}, \bibinfo {author}
  {\bibfnamefont {A.~A.}\ \bibnamefont {Tsirlin}}, \ and\ \bibinfo {author}
  {\bibfnamefont {P.}~\bibnamefont {Gegenwart}},\ }\href {\doibase
  10.1103/PhysRevLett.120.237202} {\bibfield  {journal} {\bibinfo  {journal}
  {Phys. Rev. Lett.}\ }\textbf {\bibinfo {volume} {120}},\ \bibinfo {pages}
  {237202} (\bibinfo {year} {2018})}\BibitemShut {NoStop}%
\end{thebibliography}%

\end{document}